\documentclass[twocolumn,twocolappendix]{aastex63}
\usepackage{amsmath}

\usepackage{pifont}
\newcommand{\cmark}{\ding{51}}%
\newcommand{\xmark}{\ding{55}}%

\usepackage{xcolor}

\begin{document}

\title[DREaM galaxy catalogs]{Deep Realistic Extragalactic Model (DREaM) Galaxy Catalogs: Predictions for a \emph{Roman} Ultra-Deep Field}

\correspondingauthor{Nicole E. Drakos}
\email{ndrakos@ucsc.edu}

\author[0000-0003-4761-2197]{Nicole E. Drakos}
\affiliation{Department of Astronomy and Astrophysics, University of California, Santa Cruz, 1156 High Street, Santa Cruz, CA 95064, USA }

\author[0000-0002-7460-8129]{Bruno Villasenor}
\affiliation{Department of Astronomy and Astrophysics, University of California, Santa Cruz, 1156 High Street, Santa Cruz, CA 95064, USA }

\author[0000-0002-4271-0364]{Brant E. Robertson}
\affiliation{Department of Astronomy and Astrophysics, University of California, Santa Cruz, 1156 High Street, Santa Cruz, CA 95064, USA }

\author[0000-0002-8543-761X]{Ryan Hausen}
\affiliation{Department of Computer Science and Engineering, University of California, Santa Cruz, 1156 High Street, Santa Cruz, CA 95064, USA }

\author[0000-0001-5414-5131]{Mark E. Dickinson}
\affiliation{National Optical-Infrared Astronomy Research Laboratory, 950 North Cherry Avenue, Tucson, AZ 85719, USA}

\author[0000-0001-7113-2738]{Henry C. Ferguson}
\affiliation{Space Telescope Science Institute, 3700 San Martin Drive, Baltimore, MD 21218, USA}

\author[0000-0002-0658-1243]{Steven R. Furlanetto}
\affiliation{Department of Physics and Astronomy, University of California, Los Angeles, CA 90024, USA}

\author[0000-0002-5612-3427]{Jenny E. Greene}
\affiliation{Department of Astrophysical Sciences, Princeton University, Princeton, New Jersey 08544, USA}

\author[0000-0002-6336-3293]{Piero Madau}
\affiliation{Department of Astronomy and Astrophysics, University of California, Santa Cruz, 1156 High Street, Santa Cruz, CA 95064, USA}

\author[0000-0003-3509-4855]{Alice E. Shapley}
\affiliation{Department of Physics and Astronomy, University of California, Los Angeles, 430 Portola Plaza, Los Angeles, CA 90095, USA}

\author{Daniel P. Stark}
\affiliation{Department of Astronomy, Steward Observatory, University of Arizona, 933 North Cherry Avenue, Tucson, AZ 85721, USA}

\author[0000-0003-2229-011X]{Risa H. Wechsler}
\affiliation{Department of Physics, Stanford University, 382 Via Pueblo Mall, Stanford, CA 94305, USA}
\affiliation{Kavli Institute for Particle Astrophysics and Cosmology, Stanford University, Stanford, CA 94305, USA}
\affiliation{SLAC National Accelerator Laboratory, Menlo Park, CA 94025, USA}



\begin{abstract}
In the next decade, deep galaxy surveys from telescopes such as the \emph{James Webb Space Telescope} and \emph{Roman Space Telescope} will provide transformational data sets that will greatly enhance the understanding of galaxy formation during the epoch of reionization (EoR). In this work, we present the Deep Realistic Extragalactic Model (DREaM) for creating synthetic galaxy catalogs. Our model combines dark matter simulations, subhalo abundance matching and empirical models, and includes galaxy positions, morphologies, and spectral energy distributions (SEDs). The resulting synthetic catalog extends to redshifts $z \sim12 $, and galaxy masses $\log_{10}(M/M_{\odot}) = 5 $ covering an area of $1\, \deg^2$ on the sky. We use DREaM to explore the science returns of a $1\, \deg^2$  \emph{Roman} UDF, and to provide a resource for optimizing ultra-deep survey designs. We find that a \emph{Roman} UDF to $\sim 30\,m_{\rm AB}$ will potentially detect more than $10^6$ $M_{\rm UV}<-17$ galaxies, with more than $10^4$ at redshifts $z>7$, offering an unparalleled dataset for constraining galaxy properties during the EoR. Our synthetic catalogs and simulated images are made publicly available to provide the community with a tool to prepare for upcoming data.
\end{abstract}



\section{Introduction} \label{sec:intro}

The basic picture of galaxy formation is well-established. Galaxies first form within the gravitational potential wells of dark matter halos, and continue to grow through the accretion of surrounding matter. Galaxies eventually produce enough radiation to ionize the intergalactic medium (IGM)---an epoch called reionization \citep[for a review see][]{robertson2010,stark2016}. Constraints from a variety of probes, including the cosmic microwave background \citep[CMB; e.g.][]{planck2018} and quasar absorption lines \citep[e.g.][]{becker2015}, indicate that reionization happens between $z=6$ and $z=9$. If galaxies dominate the contribution of photoionizing radiation, the cosmic star-formation rate density (CSFRD) provides a measure of the photoionizing rate. Observing high-redshift galaxies, to study galaxy formation and their role in reionization,  requires very deep imaging. 

Extragalactic ultra-deep surveys such as the Hubble Ultra-Deep Field  \citep[HUDF; e.g.][]{beckwith2006} and Hubble Frontier Fields \citep[HFF; e.g.][]{lotz2017} have detected galaxies to magnitudes $m_{\rm AB} \sim 30$ and have begun to  measure galaxy properties out to redshifts of $z \sim 10$. However, there are still many open questions at these high redshifts including the emergence of quiescent galaxies, the evolution of the UV luminosity function \citep[UVLF; e.g.][]{bouwens2021},  and the exact timeline and mechanism of cosmic reionization \citep[e.g.][]{bunker2004,finkelstein2012,robertson2015}. Upcoming telescopes, including \emph{James Webb Space Telescope (JWST)} and the \emph{Nancy Grace Roman Telescope (Roman)} will produce a large influx of data in the coming years that will greatly advance our understanding of galaxy evolution in the epoch of reionization (EoR). Given that \emph{Roman} is scheduled to launch in a few years, the purpose of this paper is to examine the science returns of an ultra-deep survey with \emph{Roman}.

The main advantage of \emph{Roman} compared to other space telescopes is its wide field of view (FOV); the \emph{Roman} Wide-Field Instrument (WFI) FOV is  more than 100x larger than \emph{HST}'s WFC3 and \emph{JWST}'s NIRCam. This large area will increase the number of detected galaxies, discover bright and rare sources, reduce cosmic variance, and probe the environment around galaxies and active galactic nuclei (AGN) at unprecedented redshifts. As outlined in \citet{koekemoer2019},  a potential Roman Ultra-Deep Field (UDF) survey could cover $\sim 1\, {\rm deg}^2$ and image to $m_{\rm AB} \sim 30$ in $\sim 600$ hours of exposure time per filter. This survey would elucidate the properties of the dominant ionizing sources at the time of reionization, allow tests for variations in the high-$z$ faint-end slope of the UVLF with environment, and likely provide the first galaxy clustering constraints at early times for faint galaxies. 

A prerequisite to understanding in detail  what a  \emph{Roman} UDF will be able to detect is accurate modeling of the expected observations. In particular, synthetic galaxy catalogs are useful for predicting the science returns of an upcoming survey, to test analysis tools, and identify potential observational biases \cite[e.g.][]{williams2018,yung2019a,korytov2019,behroozi2020,somerville2021}. To make accurate predictions, the quality and complexity of synthetic observations needs to increase with expanding theoretical and observational knowledge. This work presents the Deep Realistic Extragalactic Model (DREaM), a model for generating synthetic galaxies out to redshifts past the EoR. DREaM accurately reproduces a wide range of theoretical and observational trends, including stellar mass functions (SMFs), and the CSFRD. We use DREaM to create synthetic data for a potential $\emph{Roman}$ UDF, to provide synthetic catalogs for the community to help develop analysis and pipeline tools, and quantify the potential scientific returns of a  $\emph{Roman}$ UDF.  

To accurately capture the environment around each galaxy, we begin with dark matter simulations, and then create an observed lightcone by stitching together the discrete simulation outputs. The outline of this paper is as follows: an overview of DREaM is given in Section~\ref{sec:overview}. The underlying dark matter simulations, and the method for assigning galaxies to dark matter halos is given in Section~\ref{sec:halocat}, and the  process of creating the observed lightcone in Section~\ref{sec:lightcone}. The galaxy morphologies and spectral energy distributions (SEDs) are assigned as outlined in Sections~\ref{sec:morph} and  Section~\ref{sec:SEDs}, respectively. The resulting star formation history of the universe is discussed in Section~\ref{sec:csfh}, and preliminary predictions for the science returns of a $1 \deg^2$ UDF with \emph{Roman} are given in Section~\ref{sec:predicts}. Finally, the implications of this study and future work are discussed in Section~\ref{sec:discuss}.

\section{Overview of methods } \label{sec:overview}

While knowledge of galaxy distributions and properties comes primarily from observations, dark matter structure is typically studied through numerical simulations. The galaxy--halo connection describes how visible galaxies relate to the underlying dark matter structure \citep[for a recent review see][]{wechsler2018}. Many different approaches have been used to link halo properties to galaxies, including hydrodynamical simulations  \citep[e.g.][]{katz1991,katz1992,vogelsberger2014}, semianalytic models (SAMs) \citep[e.g.][]{white1991,kauffmann1993,somerville1999,guo2013}, halo occupation distribution models \citep[HOD models;][]{jing1998,berlind2002,wechsler2002}, abundance matching \citep[AM; e.g.][]{kravtsov2004,conroy2006,vale2006}, and machine learning \citep[e.g.][]{jo2019,moster2020,wechsler2021}. These different techniques range from physically driven, computationally expensive approaches to empirical models designed to reproduce known observational trends.

More physically based approaches are theoretically more predictive than empirical models. However, both hydrodynamical simulations and SAMs typically struggle to reproduce observed trends to the same accuracy as empirical models (which match observations by construction).\footnote{Though there has been success with the Santa Cruz SAM \citep{somerville2015} in making predictions for upcoming \emph{JWST} observations \cite[][]{yung2019a,yung2019b,yung2020a,yung2020b}.} 
Therefore, to successfully reproduce observational trends (including luminosity functions, star formation histories, and galaxy clustering), synthetic galaxy catalogs most commonly use empirical models that place galaxies in dark matter structure using AM related methods \cite[e.g.][]{moster2018,derose2019, behroozi2019, behroozi2020, derose2021} or HOD models \cite[e.g.][]{vandenbosch2005,zu2015}.

For the synthetic catalog presented in this work, we begin with dark matter simulations, and then use subhalo abundance matching (SHAM) to model the galaxy--halo connection, as outlined in Section~\ref{sec:halocat}. The main advantages to this approach are that by beginning with dark matter simulations, we can accurately capture the large-scale structure, and provide host dark matter halo properties for each galaxy. SHAM methods reproduce the proper SMFs by construction, and are known to reproduce the spatial distribution of galaxies in the local Universe \citep[e.g.][]{kravtsov2004,reddick2013,lehmann2017}.

\begin{figure}
	\includegraphics[width=\columnwidth]{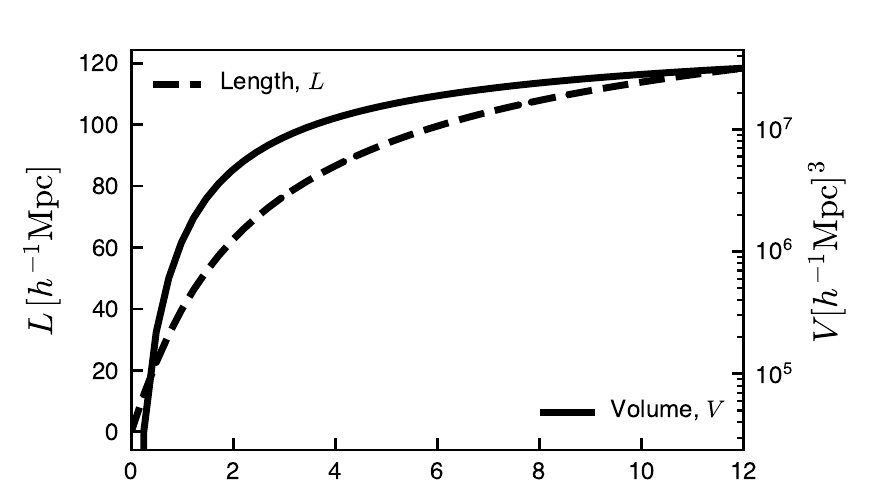}
	\caption{Comoving transverse size, $L$, and volume, $V$, of a survey with a square $1\,{\rm deg}^2$ field of view. We calculate distances assuming a Planck 2018 cosmology \citep{planck2018}. A redshift of $z=10$ corresponds to a comoving size of $\sim 115$Mpc$h^{-1}$ in the transverse direction.}
	\label{fig:dVdZ}
\end{figure}

To simulate a  $1\,{\rm deg}^2$ survey that extends past $z\sim10$, we use a simulation volume with sides of comoving length $115\,h^{-1} {\rm Mpc}$. Fig.~\ref{fig:dVdZ} shows the comoving tranverse size of a patch of sky covering $1\deg^2$ as a function of redshift, and the corresponding volume of the survey. To reach the desired depth of the synthetic realization,  we tile 60 boxes in the line-of-sight direction, as detailed in Section~\ref{sec:lightconecross}.

When constructing synthetic observations from simulations, the discrete time snapshots of the simulations need to be related to the observable sky, in which the distance of the object corresponds to its observed time. This relation between simulation data and the observable sky can be achieved by creating a lightcone. Our lightcone pipeline is described in detail in Section~\ref{sec:lightcone}, but, in brief, given merger histories for each halo we calculate if, and at which time, each halo crosses the observers past lightcone. If the position of the galaxy on the lightcone falls within the survey volume, the galaxy is included in our synthetic catalog.

After creating the lightcone, we assign galaxy morphological properties and SEDs in a manner similar to the phenomenological model from \cite{williams2018}. Details of these procedures are outlined in Sections~\ref{sec:morph}, and  Section~\ref{sec:SEDs}. We use the publicly available Flexible Stellar Population Synthesis \citep[FSPS;][]{conroy2009b,conroy2010} code to generate galaxy SEDs and calculate galaxy fluxes in each of the proposed \emph{Roman} filters. 

\begin{figure}
	\includegraphics[width=\columnwidth]{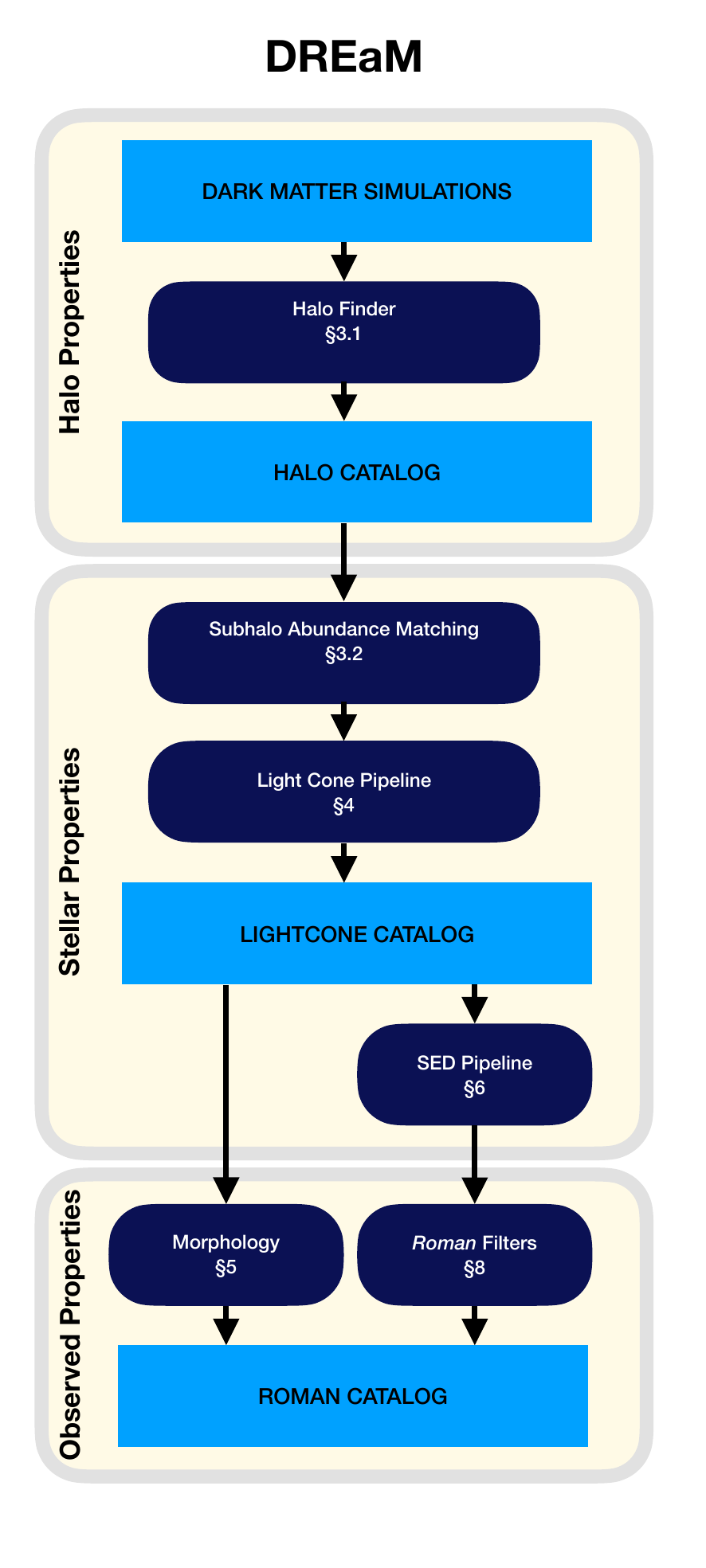}
	\caption{Overview of methods used to create the \emph{Roman} DREaM galaxy catalog, with the corresponding paper sections labeled. The galaxy catalog is based on a dark matter simulation. SHAM is used to assign galaxy masses to each dark matter halo, and then galaxy properties are assigned using empirical relations. Data products are shown with light blue rectangles, and pipelines are shown with dark blue ovals. The SED pipeline is shown in more detail in Fig.~\ref{fig:SED_Pipeline}. }
	\label{fig:Overview}
\end{figure}

A summary of the methods used to generate the synthetic catalogs is shown in Fig.~\ref{fig:Overview}. Section~\ref{sec:simulation} describes the underlying dark matter simulation and creation of the halo catalog. Sections~\ref{sec:GHC} and \ref{sec:lightcone} outline the SHAM and lightcone procedure, respectively. We assign morphologies and SEDs to the galaxies as described in Sections~\ref{sec:morph} and \ref{sec:SEDs}, respectively). Then, Section~\ref{sec:predicts} outlines the \emph{Roman} photometry.

\section{Generating synthetic galaxies} \label{sec:halocat}

As described above, we begin our catalog by running dark matter only simulations, and then use SHAM to assign stellar masses, resulting in halo catalogs with corresponding galaxy masses for every simulation time output. This section describes the underlying dark matter simulations, halo catalogs, and SHAM procedure. In the following section, we use these galaxies to create a lightcone realization of the catalog. We treat host halos and their subhalos (i.e. halos that exist inside a larger halo as a self-bound structure) separately, and use the term ``halo" to refer to both host and subhalos.

\subsection{Dark matter simulation} \label{sec:simulation}

\begin{figure*}
	\includegraphics[trim={0 1cm 0 1cm},clip]{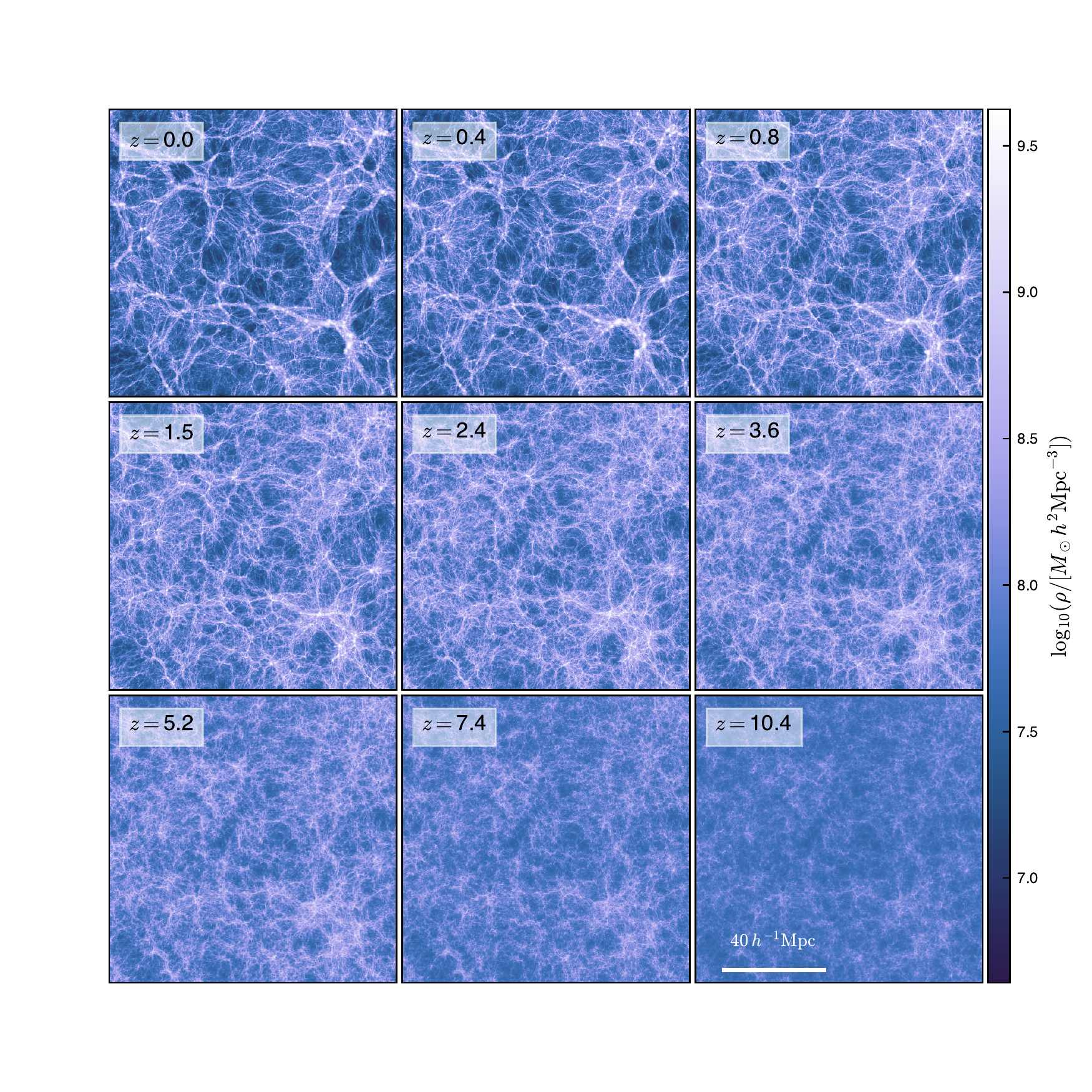}
	\caption{Cosmological $N$-body simulation with a box size of $115 h^{-1} {\rm Mpc}$. The figure shows the projected mass density at various redshifts (as labeled), in a $7 \,{\rm Mpc/h}$ slice in the radial direction. These simulations serve as the basic input for finding halos and modeling the galaxy population.}
	\label{fig:Snapshot}
\end{figure*}

We use a Planck 2018 cosmology \citep[][$\Omega_b=0.04893$, $\Omega_0=0.3111$, $\Omega_\Lambda=0.6889$,  $H_0 = 67.66\,{\rm km/s/Mpc}$, $\sigma_8= 0.8102$ and $n_s = 0.9665$]{planck2018}, with a box size of $115\,h^{-1} {\rm Mpc}$ and $N=2048^3$ particles. This corresponds to a particle mass of $1.5 \times 10^7 M_\odot h^{-1}$, The initial conditions are created using \textsc{Music} \citep{MUSIC}, and the simulations are run in \textsc{Gadget}-2 \citep{Gadget2}, with a softening length of $1.13$ comoving $h^{-1} {\rm kpc}$. The simulation outputs are shown in Fig.~\ref{fig:Snapshot}. We generate halo catalogs and merger trees using \textsc{Rockstar} \citep{rockstar} and  \textsc{Consistent Trees} \citep{consistenttrees}. We define halo masses using the  \cite{bryan1998} virial definition. Halos are required to have at least 20 particles, which corresponds to a minimum mass of $8.50 M_{\odot}h^{-1}$.

\subsection{Galaxy masses} \label{sec:GHC}

Given the halo lightcone catalog constructed in Section~\ref{sec:halocat}, we assign a galaxy stellar mass, $M_{\rm gal}$, to each halo. The SHAM procedure we use can be expressed mathematically as:
\begin{equation} \label{eq:SHAM}
\int_x^\infty n(x',z) {\rm d}x' = \int_{M_{\rm gal}}^\infty \phi (M_{\rm gal}',z) {\rm d}M_{\rm gal}'
\end{equation}
where  $\phi (M_{\rm gal},z)$ is the SMF (i.e.~the comoving number density of galaxies per unit galaxy mass, per unit redshift), and $n(x)$  is the number density of halos as a function of some mass proxy, $x$. 

In the simplest formulation of SHAM,  galaxy mass is directly mapped to halo mass. However, subhalo properties at the time of accretion are a better indicator of galaxy mass, since as subhalos undergo tidal stripping, they lose a significant amount of dark matter before the galaxy is disrupted \cite[e.g.][]{conroy2006,vale2006}. There are many different choices of mass proxies, and each of these choices produces slightly different spatial distributions of galaxies. We adopt the peak of the maximum circular velocity over the entire merger history, $V_{\rm peak}$, as the halo mass proxy, because SHAM with $V_{\rm peak}$ as the mass proxy is known to reproduce the small scale clustering of galaxies  \citep{reddick2013, hearin2013, lehmann2017, campbell2018}.

We use the SMFs from \cite{williams2018}, which are continuously evolving double-Schechter functions for both star-forming and quiescent galaxies out to redshifts $z>10$. By construction, these SMFs match observational constraints at low redshifts ($z\leq4$), and reproduce known UVLFs when convolved with the distribution of rest-frame UV magnitudes, $M_{\rm UV}$, i.e.:
\begin{equation} \label{eq:UVLF}
\begin{aligned}
\Phi_{\rm UV}(&M_{\rm UV},z) \\
&= \int_0^\infty \phi (M_{\rm gal}' )\mathcal{N} [M_{\rm UV}, \bar{M}_{\rm UV}(M_{\rm gal}', z), \sigma_{\rm UV}] {\rm d}M_{\rm gal}' \,\,\, ,
\end{aligned}
\end{equation}
where $\mathcal{N} [M_{\rm UV}, \bar{M}_{\rm UV}(M_{\rm gal}, z), \sigma_{\rm UV}]$ is a normal distribution centered on the average UV magnitude  $\bar{M}_{\rm UV}$ and $\sigma_{\rm UV}$ is the scatter in the $M_{\rm gal}$--$\bar{M}_{\rm UV}$  relation. \cite{williams2018} provide a relation to describe the $M_{\rm gal}$--$\bar{M}_{\rm UV}$ relation and scatter, which we use to model our galaxies (see Section~\ref{sec:SFGs}).

For redshifts $z \le 4$, \cite{williams2018} fit the double-Schechter
function to the data from \cite{tomczak2014}. For quiescent galaxies (QGs), above $z\sim4$ the Schechter parameters are extrapolated to higher redshifts. This extrapolation is in agreement with the few constraints for quiescent galaxies at $z>3.5$. For star-forming galaxies (SFGs), the high redshift SMFs are inferred from UVLFs. Specifically, \cite{williams2018} used data from \cite{bouwens2015} for $4 \leq z \leq 8$ and from \cite{oesch2018} at $z=10$ to fit the SMF in  Equation~\eqref{eq:UVLF}. Beyond $z\sim10$, the SMFs for the SFGs are extrapolated.

We perform SHAM on the halo catalogs using the SMFs described above. Specifically, given $N$ halos, we first find the minimum galaxy mass, $M_{\rm min}$,  such that $N_{\rm gal}(>M_{\rm min}) = N$, where
\begin{equation}
N_{\rm gal}(>M_{\rm min}) = V \int_{M_{\rm min}}^{\infty} \phi(M_{\rm gal}') {\rm d}M_{\rm gal}' \,\,\,,
\end{equation}
and $V$ is the comoving volume of the simulation. We then sample $N$ galaxy masses from the  SMF above $M_{\rm min}$. Finally, we rank-order the galaxy masses and assign galaxy masses to halos (such that the halo with the largest $V_{\rm peak}$ value gets the largest galaxy mass). 

Scatter is commonly introduced in the relation between galaxy and halo properties by either deconvolving the SMF \citep[e.g.][]{behroozi2010} or by directly adding scatter to the stellar masses, re-ranking, and iteratively solving for the galaxy mass \citep[][]{hearin2013}. Since our main goals are not to reproduce the observed scatter in the stellar-mass--halo-mass (SMHM) relation, we did not include scatter in the SHAM procedure. As illustrated in Section~\ref{sec:gallightcone}, we are able to produce realistic statistical galaxy properties, such as galaxy clustering with our approach.

We classify each galaxy as either a star-forming galaxy (SFG) or a quiescent galaxy (QG) by randomly generating a number and comparing it to the probability a galaxy of that mass is star-forming, as calculated from the SMFs. This method does ignore a possible correlation between SFR and mass accretion histories \citep[e.g.][]{behroozi2019}, which could potentially impact galaxy clustering with SFR. In particular, QGs are more likely to be found in denser environments \citep[e.g.][]{kauffmann2004,kimm2009}. Finally we remove all galaxies from the catalog with $M_{\rm gal}< 10^5  M_{\odot}$. As shown in Section~\ref{sec:predicts}, the fraction of detectable galaxies below this mass is zero for redshifts $z>1$.\footnote{Though faint high-redshift galaxies can also be detected through galaxy lensing, we expect this number to be very low for low-mass galaxies. For example, \cite{kikuchihara2020} used gravitational lensing techniques to detect very faint drop out galaxies in the HFFs to redshifts $z\sim6$, and detected galaxies with $M_{\rm gal}> 10^6  M_{\odot}$}

\section{Lightcone pipeline} \label{sec:lightcone}

Thus far we have created halo catalogs with galaxy masses at different time outputs. In this section we detail how we construct a lightcone from these discrete snapshots to create a $1 \deg^2$ survey of galaxies (the ``Lightcone Catalog"). In Section~\ref{sec:gallightcone}, we demonstrate that we are able to reproduce essential statistical properties of galaxy populations with this Lightcone Catalog.

\subsection{Lightcone crossing} \label{sec:lightconecross}

For this work, we consider a survey with a $1 \, \deg^2$ field of view. To reach redshifts $z \gtrsim 10$, we tile 60 boxes in the line-of-sight direction. This corresponds to a maximum redshift of 13 and a comoving distance of $6900\,{\rm Mpc}$.  As in \citet[e.g.][]{bernyk2016}, to avoid the replication of structures viewed by the observer, we randomly translate, reflect, and permute the axes of every tiled simulation box. We construct the halo lightcone by finding where each halo first crossed the observers past lightcone (i.e., the location at which light has had just enough time to reach the observer). Technical aspects associated with this procedure are outlined extensively in the literature \citep[e.g.][]{evrard2002, blaizot2005, kitzbichler2007, merson2013, bernyk2016, smith2017, korytov2019}. In this section, we closely follow the procedure used to generate the CosmoDC2 sky catalog for \emph{LSST} \citep{hollowed2019,korytov2019}.

We assume that the observer is at the origin of the coordinate system $(x,y,z)=(0,0,0)$. For each tiled box, we begin by placing all the host halos on the lightcone. For every host halo in snapshot $j$, we determine its position in the subsequent snapshot,  $r_{j+1}$. We extrapolate the halo position from $j$ by assuming constant velocity to find its extrapolated position in snapshot $j+1$, $r_{j+1,{\rm extrap}}$. If the halo did not have a descendant located in snapshot $j+1$, we set $r_{j+1} = r_{j+1,{\rm extrap}}$. Otherwise, we use:
\begin{equation}
r_{j+1} = r_{j+1,{\rm extrap}}  +{\rm int} \left(\dfrac{r_{j+1,{\rm extrap}} -  r_{j+1,{\rm desc}} }{L}\right) \times L \,\,\, ,
\end{equation}
where $L$ is the length of the simulation box, ${\rm int}$ represents integer, and $r_{j+1, {\rm desc}}$ is the position of the descendant in snapshot $j+1$. This approach allows the halo to cross the edge of the simulation box between snapshots $j$ and $j+1$, and thus $r_{j+1}$ might be outside the domain of the tiled box. In these cases, we also consider the scenario where the halo crosses the lightcone on the other side of the box; i.e., we apply periodic boundary conditions, such that 
\begin{equation}
\begin{aligned}
r_{j+1}^p &= r_{j+1} \,{\rm mod}\,L \\
r_{j}^p &= r_j +(r_{j+1} ^p- r_{j+1}) \,\,\, ,
\end{aligned}
\end{equation}
to allow the halo to cross the observer lightcone between positions $r_j^p$ and $r_{j+1}^p$.

Given the positions $r_j$ and $r_{j+1}$ for each host halo, we calculate the time the halo would cross the past lightcone, $t_e$ \citep[Equations 27-29 in][]{korytov2019}. If the halo crosses between the snapshot times $t_{j}$ and $t_{j+1}$, we calculate the position on the lightcone, $r_e$ from:
\begin{equation}
\begin{aligned}
v_{\rm lin}= (r_{j+1}-r_j)/(t_{j+1}-t_j) \\
r_e = r_j + v_{\rm lin}(t_e-t_j) \,\,\, .
\end{aligned}
\end{equation}
We do not include any halos where $r_e$ is beyond the domain of the tiled box. 

The final step to generating the lightcone is to assign halo properties (e.g.~mass and substructure) to each object. One approach to assigning halo properties is to allow the merger of halos to happen at a time randomly between the two snapshot \citep[e.g.][]{smith2017}. However, this approach will result in halos being double counted, and the lightcone needs to be carefully pruned. Since our simulations have very fine time resolution (500 snapshots between redshifts $z=0$ and $z=20$), we can avoid the technical difficulties and assumptions introduced by needing to prune the catalog, and follow the same approach from \cite{korytov2019}. Specifically, every halo that crosses the lightcone between snapshots $j$ and $j+1$ is assigned properties from snapshot $j$. When including the substructure, we ensure subhalos have the same position and velocity offset from the host as in snapshot $j$. 

\subsection{Survey volume} \label{sec:surveyvolume}

Given the lightcone catalog consisting of halo properties, galaxy masses, and positions, we cut out a wedge in the survey volume corresponding to $1\,{\rm deg}^2$. Following \citet{bernyk2016}, we convert the comoving $(x,y,z)$ positions of each galaxy to an angular position as follows:
\begin{equation}
\begin{gathered}
d = \sqrt{x^2 + y^2 + z^2}\\
{\rm RA} = \dfrac{360}{2 \pi} \arctan(y/x)\\
{\rm Dec} = \dfrac{360}{2 \pi} \arcsin(z/d)
\end{gathered}
\end{equation}
We only consider galaxies with ${\rm RA}<1\,{\rm deg}$ and ${\rm Dec}<1\,{\rm deg}$, and then center the galaxies on $({\rm RA}, {\rm Dec})=(0,0)$.

\begin{figure*}
	\includegraphics{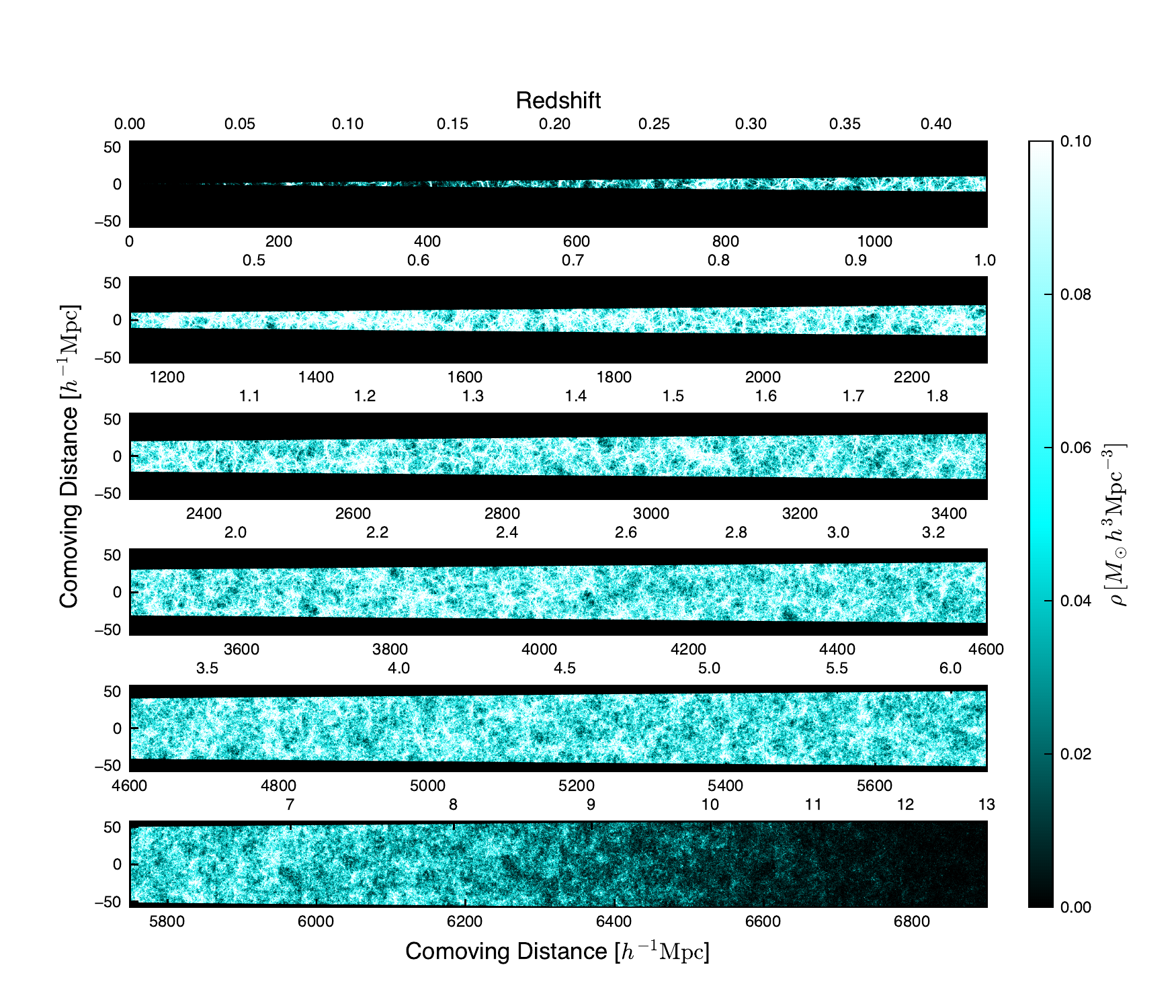}
	\caption{Mass density of galaxies in survey volume, with a field of view of $1 \, {\rm deg}^2$. The $x$-axis shows the cosmological redshift (top) and the comoving distance (bottom). We have plotted 10 tiled boxes in each row, for a total of 60 boxes to reach the desired redshift range. This figure demonstrates the depth and distribution of galaxies  of a $1\,\deg^2$ UDF.}
	\label{fig:HaloLightCone}
\end{figure*}

Fig.~\ref{fig:HaloLightCone} shows the resulting galaxy mass density as a function of position. The lightcone procedure results in a realistic distribution of galaxies that trace the underlying cosmic web. Since the survey will reach very deep distances, the survey is tiled in 6 rows, with each row showing 10 tiled boxes. The survey wedge is complete to a maximum distance of $6588\, \rm{Mpc/h}$ (which corresponds to a cosmological redshift of $\sim10.5$); at higher reshifts, the angle of the survey is larger than the width of the simulation box. There are discontinuities in the galaxy density where the simulation boxes are tiled, which is a common feature for lightcones. As addressed in \citet{bernyk2016}, statistical properties of the galaxy catalog will be accurate on scales smaller than the box size.

\subsection{Statistical properties of galaxy lightcone} \label{sec:gallightcone}

To verify the lightcone pipeline results in a realistic galaxy population, we calculate the halo mass functions (HMFs), SMFs, stellar-to-halo mass relations (SHMRs) and galaxy clustering of the synthetic galaxy catalog.

\subsubsection{Halo mass functions}

\begin{figure}
	\includegraphics[width=\columnwidth]{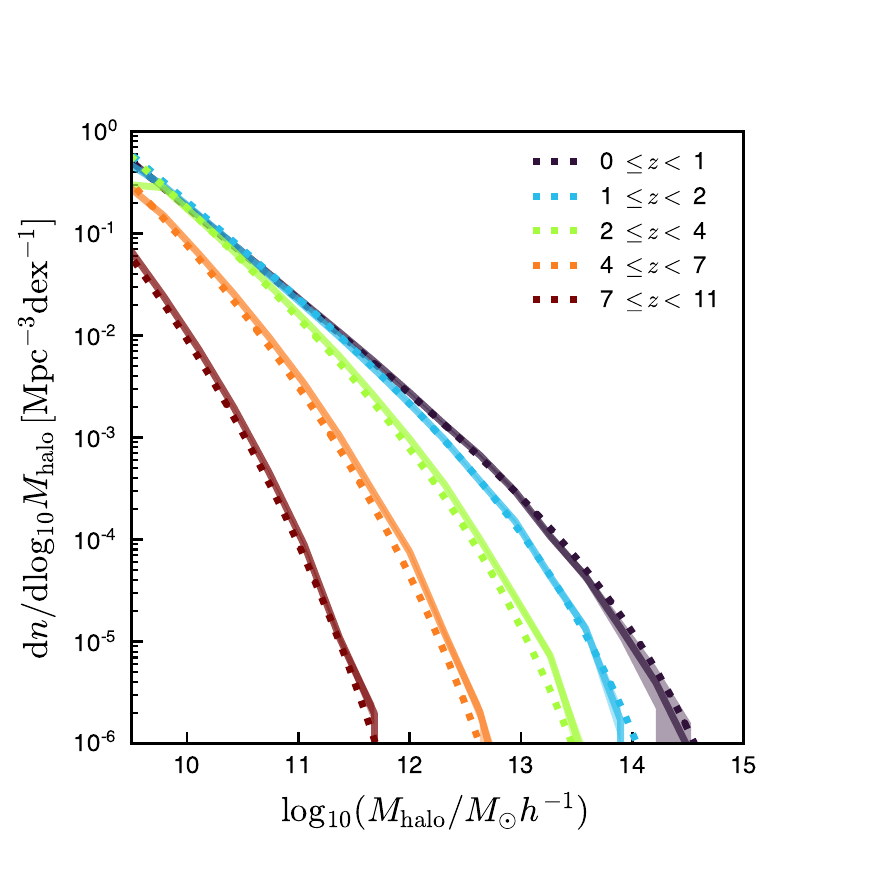}
	\caption{Expected HMF from the \citet{despali2016} parameterization (dotted lines) compared to the host halos in the lightcone catalog (solid lines). The theoretical curve is the volume-weighted average over the redshift band. The lightcone catalog agrees with the expected HMF, verifying that the lightcone procedure reproduces the correct redshift and mass distribution of dark matter halos.}
	\label{fig:HMF_lightcone}
\end{figure}

To demonstrate the HMF of the halo lightcone catalog is correct, we plot the recovered HMF of host halos in Fig.~\ref{fig:HMF_lightcone}. The theoretical curve is from the parameterization from  \cite{despali2016}, and calculated using the HMF routine in \textsc{Colossus}  \citep{colossus}. The theoretical curve and simulation results agree very well. The catalog was constructed to be complete above galaxy masses $M_{\rm gal} = 10^5 M_\odot$, which corresponds to a halo mass limit of approximately $M_{\rm halo} \sim10^{9.5} h^{-1} M_\odot$

\subsubsection{Stellar mass functions and the stellar-to-halo mass relations}

\begin{figure}
	\includegraphics[width=\columnwidth]{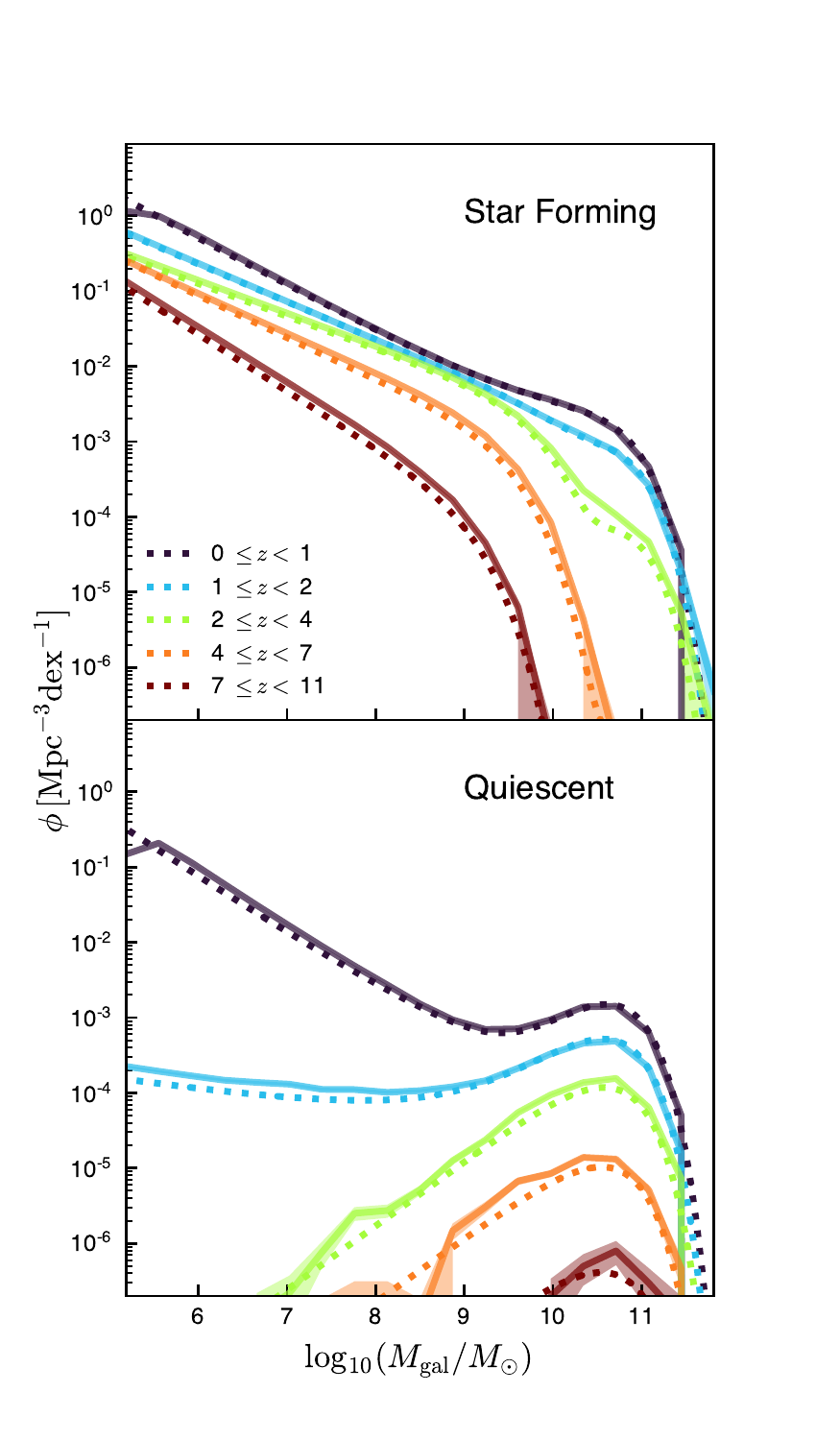}
	\caption{SMFs for the lightcone catalog. SFGs and QGs are shown in the top and bottom panels, respectively. The  \cite{williams2018} SMFs that were used in the abundance matching procedure are shown with dotted lines, calculated as a volume-weighted average over the redshift bin. The lightcone pipeline and abundance matching procedure reproduce the desired SMFs.}
	\label{fig:AbundanceMatching}
\end{figure}

Additionally, we show the SMF of the lightcone catalog in Fig.~\ref{fig:AbundanceMatching} for  both SFGs and QGs. For comparison, we show the  volume-weighted average SMFs from \cite{williams2018} (dotted lines), averaged over the redshift bin. The catalog is complete at all redshifts for galaxy masses greater than $10^5 \,M_{\odot}$. The galaxies in the lightcone agree with the SMFs from \cite{williams2018}, which verifies that the lightcone pipeline and abundance matching procedure reproduce realistic galaxy counts.   As discussed in Section~\ref{sec:GHC}, producing the desired SMF is important for reproducing observed luminosity functions.

\subsubsection{Galaxy clustering}

Galaxy clustering is commonly described using the two-point correlation function (2PCF), $\xi(r)$, defined as
\begin{equation}
{\rm d} P  = n_g (r) [1+ \xi(r)] {\rm d}V \,\,\, ,
\end{equation}
where ${\rm d}P$  is the excess probability above Poisson noise, $n_g(r)$ is the mean density of galaxies at separation $r$, and ${\rm d} V$ is the differential volume. In practice, the projected 2PCF, $w_p$, is often used. The 2PCF depends on the projected separation, $r_p$, and the line-of-sight separation, $\pi$:
\begin{equation}
w_p = \int_{-\infty}^{\infty} \xi (r_p,\pi) {\rm d} \pi \approx 2 \times \int_{0}^{\pi_{\rm max}} \xi (r_p,\pi) {\rm d} \pi \,\,\, ,
\end{equation}
where the second equality assumes isotropy. The upper limit, $\pi_{\rm max}$, needs to be chosen large enough to nullify the effect of peculiar velocities, but not so large that it creates artificial edge effects from the survey boundary. The optimal value for  $\pi_{\rm max}$ depends on the underlying survey volume, but is typically in the range 40--80 Mpc \citep{zehavi2005,vandenbosch2013}.

We calculate $\xi$ based on the commonly used Landy-Szalay estimator \citep{landy1993}, 
\begin{equation}
\xi \approx \dfrac{DD-2DR + RR}{RR} \,\,\, ,
\end{equation}
where $DD$, $DR$ and $RR$ are normalized number of counts from data--data, data--random and random--random pairs. 
The random catalog contains $10^3$ points per ${\rm arcmin}^2$, with distances assigned such that the random catalog has the same redshift distribution as the synthetic catalog, i.e.,
\begin{equation}
\begin{aligned}
N&(z_{\rm min}<z<z_{\rm max}) \\
&=\theta \sin(\theta) \int_{\chi(z_{\rm min})} ^{\chi(z_{\rm max})} \chi^2  \int_{M_1}^{M_2}  \phi(M,z(\chi)) {\rm d} M {\rm d} \chi
\end{aligned}
\end{equation}
where $\theta$ is the survey angle, and $\phi$ is the  \cite{williams2018} SMF. Values for RA and Dec coordinates for each random galaxy were selected so that galaxies are distributed isotropically:
\begin{equation}
\begin{aligned}
{\rm RA} &= r_1  \theta  \\
{\rm Dec} &=  \arcsin \left[ 2r_2 \sin\left(\dfrac{\theta}{2}\right) \right] \,\,\, .
\end{aligned}
\end{equation}
where $r_1$, $r_2$ are random numbers selected uniformly between -0.5 and 0.5.

\begin{figure}
	\includegraphics[width=\columnwidth]{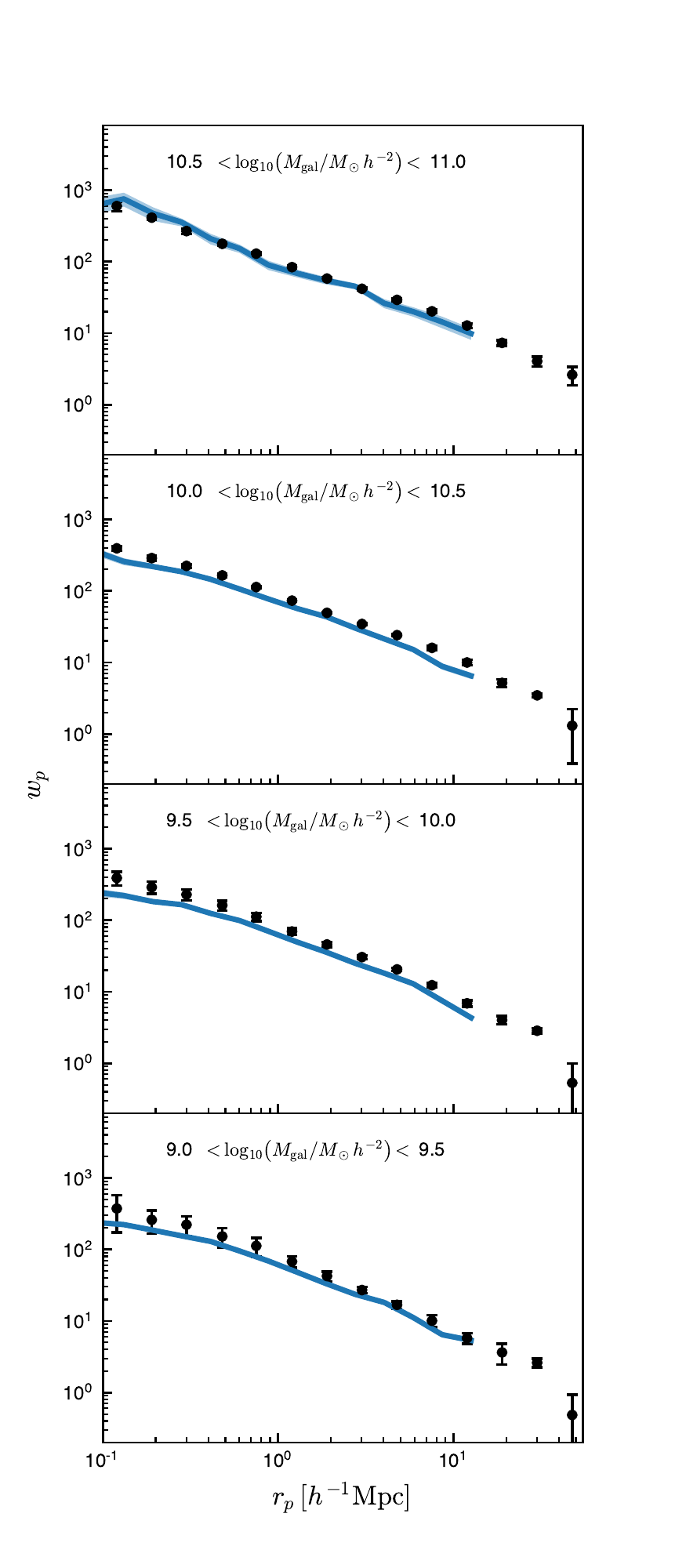}
	\caption{2PCF in a $100 \deg^2$ lightcone catalog for all galaxies $0.1<z<0.2$ (lines). We estimate uncertainties in the synthetic catalog 2PCF  by bootstrapping the data. In comparison, SDSS data from \citet{yang2012}, for galaxies at redshifts $z \approx 0.1$ are shown with black points. The agreement between the simulation and data shows that the lightcone catalog matches the observed galaxy clustering at low redshifts.}
	\label{fig:Clustering_lightcone}
\end{figure}

To measure the projected 2PCF we use the package \textsc{Corrfunc} \citep{sinha2020}, with $\pi_{\rm max}= 60 \, h^{-1}\, {\rm Mpc} $. We estimateerrors in the measured 2PCF by bootstrapping the galaxy catalog data with 200 subsamples. We present this measured 2PCF in Fig.~\ref{fig:Clustering_lightcone}, along with SDSS data for redshift $z\approx0.1$ galaxies from \citet{yang2012}. 
To compare our synthetic catalog 2PCF to the data, we use galaxies in the redshift range $0.1<z<0.2$ within a $10 \deg \times 10 \deg$ survey. The low redshift clustering of the synthetic galaxies agree with observations within the $1-\sigma$ error bars, indicating that the lightcone halo clustering is realistic.

\section{Galaxy morphologies} \label{sec:morph}

Galaxy morphologies offer key insights into galaxy evolution, and are important for understanding observational systematics. We model all galaxies as \cite{sersic1968} profiles with an index, $n_s$. Additionally, all galaxies are assigned a projected size $R_{\rm eff}$, a projected axis ratio $q=b/a$, and a position angle (PA). We follow the morphological prescriptions from  \cite{williams2018}---which are based on the empirical distributions measured in \emph{HST} images---with the exception of galaxy sizes. 

\subsection{Size}

Galaxies sizes are known to decrease with increasing redshift at a fixed stellar mass \citep[e.g.][]{vanderwel2014,shibuya2015, curtislake2016}, typically evolving as
\begin{equation}
R_e \propto (1+z)^{-\alpha} \,\,\, ,
\end{equation}
where $R_e$ is the half-light radius. Further, SFGs and QGs evolve differently, with  star-forming galaxies having lower values of $\alpha$ \citep[e.g.][]{vanderwel2014, ma2018}.\footnote{However, \cite{miller2019} recently  showed that differences between size trends for SFGs and QGs may disappear if you use $r_{80}$---the radius enclosing 80 \% of stellar light---rather than $R_e$.} In addition to these observed trends, theoretical models commonly show a correlation between the size of a galaxy and that of its host dark matter halo \cite[e.g.][]{Mo1998,kravstov2013,jiang2019}. Given that our synthetic catalog begins with the underlying dark matter structure, we assign galaxy sizes based on the dark matter halo sizes, and then test that the size--redshift--mass relation is consistent with current data.

Specifically, we characterize the size of each galaxy as the half-light radius in the semi-major axis $R_{\rm eff}$.\footnote{$R_{\rm eff}$ is related to the commonly used circularized half-mass radius, $R_{\rm eff, circ}$, by $R_{\rm eff, circ} = \sqrt{b/a} R_{\rm eff}$, where $b/a$ is the minor-to-major axis ratio.} 
To assign $R_{\rm eff}$ to each galaxy, we use the relation between halo and galaxy size \citep[e.g.][]{kravstov2013,zanisi2020}:
\begin{equation} \label{eq:Reff}
R_{\rm eff} = A \,R_{\rm vir},
\end{equation}
where $R_{\rm vir}$ is the radius of the halo in physical units. We assign $R_{\rm eff}$ values from this mean relation, with lognormal scatter, $\sigma_{R}$.  We use the coefficients $A$ and scatter $\sigma_{R}$ from \cite{zanisi2020}, as summarized in Table~\ref{tab:zanisi}. \cite{zanisi2020} used size distributions of central galaxies from the Sloan Digital Sky Survey (SDSS) DR7 sample \citep{abazajian2009,meert2016}. While these coefficients capture the dependence of $R_{\rm eff}$ on  mass and on whether a galaxy is star-forming or quiescent, they are calibrated for $z=0$ galaxies.

\begin{table}
   \caption{Coefficients, $A$, for the  $R_{\rm eff}$--$R_{\rm vir}$ relation (Equation~\ref{eq:Reff}) and the scatter $\sigma_{R}$; values are from \cite{zanisi2020}.}  \label{tab:zanisi}
   \centering
  \begin{tabular}{c c c c} 
  \tableline
    $\log_{10}(M_{\rm gal}/M_\odot)$ & $A$ (SFGs) & $A$ (QGs) & $\sigma_{R}$\\
 \tableline
    $ \leq 9.5$ & 0.018  &0.006 & 0.2\\
    $(9.5,10] $ & 0.019  &0.007& 0.2\\
    $(10,10.5] $ &0.019  & 0.010& 0.15\\
    $(10.5,11] $ &0.019  &0.011& 0.15\\
     $(11,11.5] $ & 0.019  &0.015& 0.15\\
    $ >11.5 $ & 0.024  & 0.016 &  0.1 \\
 \tableline
  \end{tabular}
\end{table}

\begin{figure*}
	\includegraphics{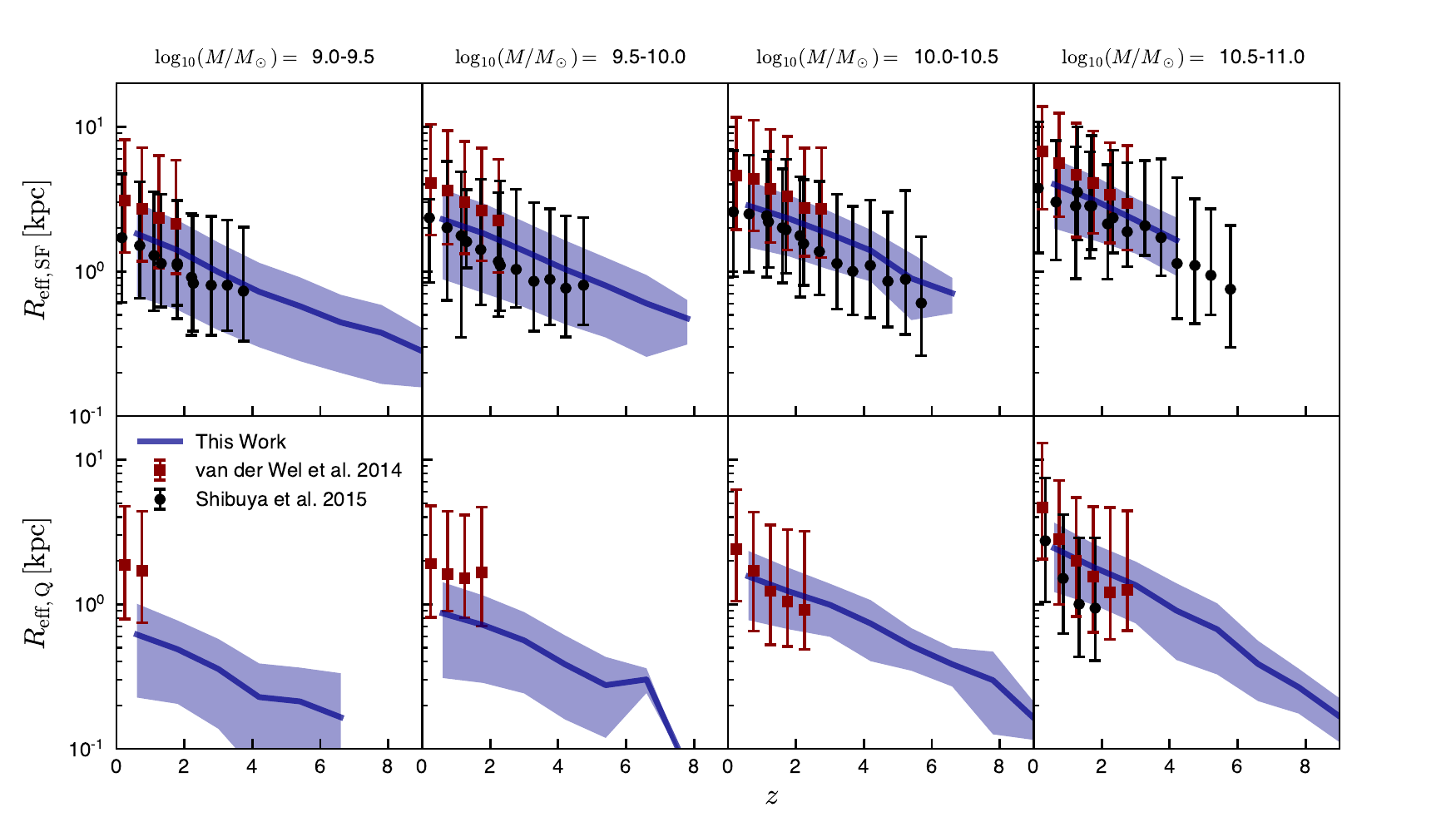}
	\caption{Median half-light semi-major radius, $R_{\rm eff}$ as a function of redshift from the synthetic galaxy catalog for SFGs (top) and QGs (bottom). Different columns show different mass bins. The shaded regions are the $1-\sigma$ standard deviation in each bin for the synthetic galaxies. The points show median data $R_{\rm eff}$ from  \cite{vanderwel2014,shibuya2015}, with the 16th and 84th percentiles of the data point distribution. Our synthetic galaxies match observed trends of galaxy sizes with mass, redshift, and galaxy classification.}
	\label{fig:Reff}
\end{figure*}

To test whether this relation  extends reasonably to higher redshifts, we compare the median $R_{\rm eff}$--$z$ relation from our catalog to data from \cite{vanderwel2014} and \cite{shibuya2015}, as shown in Fig.~\ref{fig:Reff}. 
The datasets shown in Fig.~\ref{fig:Reff}  consist of galaxies from the 3D-\emph{HST}+CANDELS catalog. Redshifts and masses are from \cite{skelton2014}, and galaxies are classified as either star-forming or quiescent using UVJ diagram criteria. 

Fig.~\ref{fig:Reff} demonstrates that our simple assumption that Equation~\eqref{eq:Reff} holds for all redshifts agrees remarkably well with the data. We capture the decrease in $R_{\rm eff}$ with both decreasing mass and increasing redshift. We also capture the effect that QGs are smaller than SFGs on average.  Our ability to capture redshift trends, despite only using known relations between $R_{\rm eff}$ and $R_{\rm vir}$ at $z=0$, likely reflects the fact that virial radii evolve as $R_{\rm vir} \propto1/(1+z)$, and therefore higher redshift galaxies have smaller virial radii. Our finding that the redshift evolution in the mass-size relation can be captured through the redshift evolution $R_{\rm vir}$ has also been found by  \cite{mowla2019b}, who show that the ratio $r_{80}/R_{\rm vir}$ is constant for redshifts $z<3$.

\subsection{Axis ratios,  Sérsic indices and Position Angles}

The distribution of galaxy axis ratios and Sérsic indices should differ between SFGs and QGs \cite[e.g.][]{franx2008,bell2012,mortlock2013} and with redshift \cite[e.g.][]{vanderwel2011,guo2015}. To assign shape (defined as the projected axis ratio $q=b/a$,  where $a$ is the semi-major half-light size and $b$ is the semi-minor half-light size) and Sérsic indices, $n_s$, we use the method directly from \cite{williams2018}. Specifically, using data compiled from \cite{vanderwel2012} and \cite{skelton2014}, \cite{williams2018} finds the distribution of   $q$ and $n_s$ as a function fo redshift. Due to limited data, morphologies for QGs with $z>4$ are drawn from the $3 \leq z \leq 4$ distribution, and morphologies for SFGs with $z>6$ are drawn from the $5 \leq z \leq 6$ distributions. We use these resulting distributions to draw $q$ and $n_s$ values to assign to each galaxy.

We assign PAs uniformly between $0$ and $2 \pi$. Synthetic galaxy catalogs often assume that galaxies are oriented isotropically \citep[e.g.][]{williams2018}, and this assumption will not affect the galaxy clustering, number counts, or cosmic star formation history of our synthetic catalog.

\section{Galaxy SEDs} \label{sec:SEDs}

In this section we outline the methods we use to generate galaxy SEDs. The SED pipeline aims to reproduce observed star formation histories and UVLFs, as these quantities dictate the number of ionizing photons produced by galaxies (see Section~\ref{sec:csfh}). The SEDs also have realistic SFG and QG properties, including their observed colors, ages, and metallicity (see Appendix~\ref{sec:galprops}).

\subsection{Overview} \label{sec:SED_overview}

One of the main observables included in the synthetic catalog is the photometry in the \emph{Roman} filters, which requires accurate modeling of the galaxy SEDs. We model the SEDs for each galaxy using the software FSPS \citep{conroy2009b,conroy2010}. This section serves as an overview of the model used to generate the galaxy SEDs.

We use a \cite{chabrier2003} initial mass function (IMF), and include the IGM absorption model from \cite{madau1995}.  The star-formation history model is described in Section~\ref{sec:SFH}. Additionally, for SFGs, we include the FSPS nebular emission model \citep{byler2017} which is controlled by the gas ionization parameter and the gas metallicity. As in \cite{williams2018}, we approximate the stellar and ISM metallicities as a single metallicity value, $Z_{\rm met}$.

The dust modeling includes dust emission and dust absorption. Dust emission follows \cite{draine2007}, which is a silicate-graphite-PAH grain model. The general FSPS dust absorption model follows the commonly used \cite{charlot2000} model. We use the \cite{calzetti2000} attenuation curve, where dust attenuation is applied to all starlight equally, and therefore depends on one parameter, the total attenuation optical depth $\bar{\tau}_v$. We also include the asymptotic giant branch (AGB) circumstellar dust models from \cite{villaume2015}. Including circumstellar dust can make a significant contribution to the IR emission from galaxies with little diffuse gas, and will need to be included in stellar modeling to accurately interpret data from upcoming IR facilities, including \emph{JWST}.

Overall, this leaves seven free parameters to describe the galaxy SEDs:
\begin{itemize}
\item $M_{\rm gal}$: the stellar mass of the galaxy.
\item $z$: the redshift of the galaxy (or alternatively, $t_{\rm age}$, the age of the universe at that redshift).
\item  $Z_{\rm met}$: the galaxy metallicity. The gas-phase and 
stellar metallicities are assumed to be equal.
\item $t_{\rm start}$: the age of the universe at the start of star formation.
\item $\tau$: the $e$-folding time for star-formation.
\item $\bar{\tau}_v$:  the dust attenuation parameter, defined as the opacity at $5500\AA$. 
\item $U_S$:  the gas ionization parameter.
\end{itemize}
In addition to these seven parameters, other relevant quantities included in the synthetic catalog include the star-formation rate (SFR), the UV magnitude, $M_{\rm UV}$, the slope of the UV continuum, $\beta$, and the rest-frame colors of the galaxies. Section~\ref{sec:calculated} describes the calculation of each of these quantities.

\begin{figure}
	\includegraphics[width=\columnwidth]{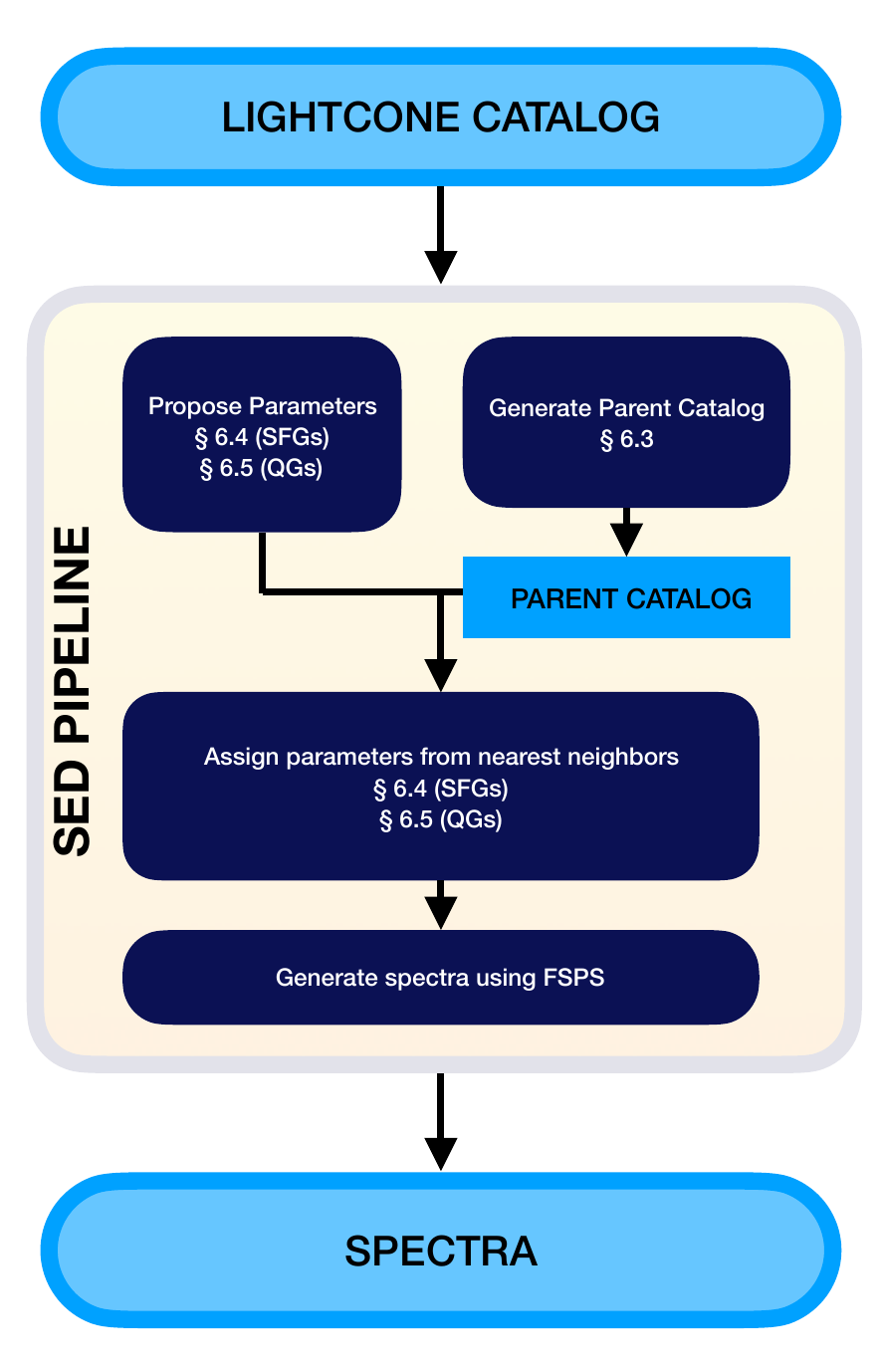}
	\caption{Overview of the SED pipeline. We begin with a galaxy catalog consisting of redshift and masses for each galaxy, and parent catalogs for the SFGs and QGs. We propose parameters for every galaxy in the catalog and find the nearest neighbors in the parent catalogs. We then assign FSPS parameters based on a weighted average of the nearest neighbors.}
	\label{fig:SED_Pipeline}
\end{figure}

Each galaxy begins with a redshift $z$ (from the lightcone pipeline), a mass, $M_{\rm gal}$ (from abundance matching), and is labeled as either a SFG or a QG, as sampled from the SMF (see Section~\ref{sec:GHC}). Sections~\ref{sec:SFGs} and \ref{sec:QGs} below describe SED pipeline for assigning the remaining FSPS parameters ($Z_{\rm met}$, $t_{\rm start}$, $\tau$, $\bar{\tau}_v$ and $U_S$)  for SFGs and QGs, respectively. The pipeline is summarized in Fig.~\ref{fig:SED_Pipeline}. 

In brief, we first generate ``parent catalogs" for both SFGs and QGs, spanning a realistic range of parameters. These parent catalogs serve as a lookup table for realistic galaxy SEDs. Details on how the parent catalogs are calculated are given in Section~\ref{sec:parent}. We propose parameters for each galaxy, and find the 10 nearest neighbors in the parent catalog for each set of galaxy parameters using k-d trees\footnote{k-d trees are $k$-dimensional data structures arranged as a binary tree. This structure  allows for rapid searches to find the nearest neighbors of a given point in the $k$-dimensional space.}. For SFGs, the proposed parameters are $M_{\rm gal}$, $z$, $M_{\rm UV}$ and $\beta$ while for QGs, the proposed parameters are  $M_{\rm gal}$ and $z$. The distance metrics used to find the nearest neighbours are detailed in Sections~\ref{sec:SFGs} and \ref{sec:QGs}.

We assign the five free FSPS parameters ($Z_{\rm met}$, $t_{\rm start}$,  $\tau$, $\bar{\tau}_v$  and  $U_S$) to each galaxy by taking the weighted average of the nearest neighbors. Given these five parameters, along with the mass and redshift of the galaxy, we use FSPS to generate an SED. We then calculate the UV properties, SFRs and rest-frame colors directly from the spectrum. These calculated properties will not be identical to the proposed parameters, however we still retain the imposed scaling relations in the parent catalogs (see Appendix~\ref{sec:galprops}).

\subsection{Calculated quantities} \label{sec:calculated}

For the SED pipeline outlined above, we need to accurately measure the SFR, rest-frame colors, and UV properties of the galaxy SEDs. This section details the methods we use to measure these quantities. Since FSPS normalizes the stellar modeling, such that one stellar mass is created over the formation history, we scale all the spectra. Specifically, we scale all fluxes and the SFR by $M_{\rm gal}/f$, where $f$ is the surviving mass fraction in the stellar population, not including stellar remnants.

\subsubsection{Star formation history} \label{sec:SFH}

Galaxy SFRs dictate the cosmic star formation rate density (CSFRD) of the universe, which in turn constrains the amount of ionizing photons produced by galaxies in the EoR. Given the importance of SFRs to galaxy evolution and the sources of ionizing photons during the EoR, we need to ensure our synthetic galaxies have SFRs consistent with observations.

The SFR depends on the age, $e$-folding time and mass of the galaxy. We model the star formation history, $\psi(t)$, using a ``delayed tau model", in which:
\begin{equation}
\psi(t) \propto t \exp{-t/\tau} \,\,\, ,
\end{equation}
where $t$ is the time since the start of star formation, $t_{\rm start}$. As discussed in \cite{williams2018}, this parameterization achieves the expectation from simulations that high-redshift galaxies have rising star formation histories \citep{finlator2011}, and accurately reproduces the colors and mass-to-light ratios of galaxies in smoothed particle hydrodynamics (SPH) simulations  \citep[e.g.][]{simha2014}. 

To determine the SFR for each galaxy, we use the FSPS calculated SFR, and scale it by  $M_{\rm gal}/f$, as described above.

\subsubsection{Rest-frame colors} \label{sec:UVJ}

UVJ diagram galaxy classifications allow for the separation of QGs and dusty SFGs  \cite[e.g.][]{williams2009,whitaker2013,papovich2018}. To examine the distribution of the synthetic galaxies in  UVJ space, we use calculated U-V and V-J colors. Specifically, we use the Johnson U and V filters, and the  WFC3 F125W J filter included in FSPS.

\subsubsection{UV properties} \label{sec:UVproperties}

\emph{Roman} will constrain the number of ionizing photons that are produced by galaxies in the EOR by precisely measuring the UVLF at high redshift. The UVLF depends on the SMF and the UV magnitude, $M_{\rm UV}$ (see Equation~\ref{eq:UVLF}). 

Following \cite{robertson2013} we define the UV magnitude, $M_{\rm UV}$, as the average magnitude at rest-frame wavelength in a flat filter, in the range 1450-1550\AA. We measure $M_{\rm UV}$ directly from the rest-frame galaxy SEDs by calculating the average flux density:
 \begin{equation}
 \langle f_\lambda \rangle =  \dfrac{\int_{\lambda_1}^{\lambda_2} f_{\lambda}d\lambda}{\lambda_2-\lambda_1} \,\,\,,
 \end{equation}
 with $\lambda_1=1450$\AA  and  $\lambda_2=1550$\AA.

In addition to the UV magnitude, the slope of the UV continuum, $\beta$, helps determine the role of galaxies in ionizing the universe. The slope $\beta$ is defined as $f_{\lambda} \propto \lambda^\beta$ \citep[e.g.][]{meurer1999}. 
The presence of very blue UV continuum slopes in Roman imaging may provide a signpost of very high escape fractions in early galaxies. We use the method from \cite{dunlop2012} to determine $\beta$. Specifically, we take the SEDs shifted to redshift $z=7$, and calculate  
\begin{equation}
\beta = 4.43(J125-H160)-2 
\end{equation}
using the FSPS included filters F125W and F160W for the Wide Field Camera 3 (WFC3).

\subsection{Parent Catalogs} \label{sec:parent}

We create the parent catalogs by sampling from known scaling relations. By only populating regions of parameter space corresponding to the desired galaxy population, we can fully sample the relevant space, while also limiting the size of the parent catalog. The distributions we use to construct the parent catalogs are summarized in Tables~\ref{tab:SFGs} and \ref{tab:QGs} for SFGs and QGs, respectively, and outlined in detail in this section.

\begin{table*}
  \centering
   \caption{Parameters for SEDs of SFGs. The synthetic galaxies have a fixed parameters $M_{\rm gal}$ and $z$ and ``proposed"  parameters  $M_{\rm UV}$ and $\beta$. We identify the 10 closest neighbours in the parent catalog (defined by the distance metric), and assign the final catalog properties $Z_{\rm met}$, $U_S$, $\bar{\tau}_V$, $t_{\rm start}$ and $\tau$ from the nearest neighbors.}  \label{tab:SFGs}
  \begin{tabular}{c  c c c c} 
  \tableline
     Parameter &  Proposed Parameter & Parent Catalog &Distance Metric & Assigned Parameter\\
 \tableline
   $M_{\rm gal}$ 
	& fixed
   	& uniform in $\log_{10}(M_{\rm gal}/M_\odot)\in[5,12]$
	& $\log_{10}(M_{\rm gal}/10 M_\odot)$
	& fixed \\
   $z$ 
	& fixed 
   	& uniform in $a=1/(1+z)\in[0.07,1]$ 
	&  $a=10/(1+z)$ 
	& fixed \\
\hline
   $Z_{\rm met}$ 
   	& ... 
    	& FMR 
  	& ...
  	 & nearest neighbors \\
   $U_S$
   	 & ... 
	 & $U_S$--$Z_{\rm met}$   relation
	 & ...
	 & nearest neighbors\\
   $\bar{\tau}_V$ 
	& ...
   	& $\mathcal{N}(\mu=0, \sigma=0.5)\in[0,4]$
	& ... 
	& nearest neighbors\\
   $t_{\rm start}$ 
	& ...
   	&  uniform in $[1\,{\rm Myr},t_{\rm age}]$
	& ...
	& nearest neighbors \\
   $\tau$ 
	& ...
   	& uniform in $[0.1,100]$ Gyr
	& ...
	& nearest neighbors \\
\hline	
   $\psi$ 
	&  ...
   	& calculated; cut $\log_{\rm 10} (\psi/M_\odot {\rm yr}^{-1})\in[-5,4]$
	& ...
	& calculated \\
   $M_{\rm UV}$
	& $M_{\rm UV}$--$M_{\rm gal}$
   	& calculated; cut in $[-25,-5]$
	& $M_{\rm UV}$ + 20
	& calculated\\
   $\beta$ 
	& $\beta$--$M_{\rm UV}$
   	& calculated; cut in $[-3,-1]$
	& $\beta$
	& calculated\\
   $U$-$V$ 
	& ...
   	& calculated
	& ...
	& calculated\\
   $V$-$J$ 
	& ...
   	& calculated
	& ...
	& calculated\\	
 \tableline
  \end{tabular}
\end{table*}

\begin{table*}
  \centering
   \caption{Parameters for SEDs of QGs. The synthetic galaxies have a fixed parameters $M_{\rm gal}$ and $z$. We identify the 10 closest neighbours in the parent catalog (defined by the distance metric), and assign the final catalog properties $Z_{\rm met}$, $\bar{\tau}_V$, $t_{\rm start}$ and $\tau$ from the nearest neighbors.}  \label{tab:QGs}
  \begin{tabular}{c  c c c c} 
  \tableline
     Parameter  & Proposed Parameter & Parent Catalog & Distance Metric & Assigned Parameter\\
 \tableline
   $M_{\rm gal}$ 
	& fixed
   	& uniform in $\log_{10}(M_{\rm gal}/M_\odot)\in[5,12]$
	& $\log_{10}(M_{\rm gal}/10 M_\odot)$
	& fixed \\
   $z$ 
	& fixed 
   	& uniform in $[0,13]$ 
	&  $a=10/(1+z)$ 
	& fixed \\
\hline
   $Z_{\rm met}$ 
   	& ... 
   	&  FMR 
   	& ... 
  	 & nearest neighbors \\
   $\bar{\tau}_V$ 
	& ...
   	& $\mathcal{N}(\mu=0, \sigma=0.5)\in[0,4]$
	& ... 
	& nearest neighbors\\
   $t_{\rm start}$ 
	& ...
   	&  uniform in $[1\,{\rm Myr},t_{\rm age}]$
	& ...
	& nearest neighbors \\
   $\tau$ 
	& ...
   	& uniform in $[0.01,10]$ Gyr
	& ...
	& nearest neighbors \\
\hline	
   $\psi$ 
	&    ...
   	& calculated; cut $\log_{10} (\psi/M_\odot {\rm yr}^{-1})\in[-4,1]$
	& ...
	& calculated \\
   $U$-$V$ 
	& ...
   	& calculated; cut in UVJ diagram
	& ...
	& calculated\\
   $V$-$J$ 
	& ...
   	& calculated; cut in UVJ diagram
	& ...
	& calculated\\	
 \tableline
  \end{tabular}
\end{table*}

\subsubsection{Star-forming galaxy parent catalog}

For the SFG parent catalog, we sample the seven FSPS parameters for $10^6$  galaxies. The mass, redshift, and SFH parameters $t_{\rm start}$ and $\tau$ are sampled from uniform distributions, as summarized in Table~\ref{tab:SFGs}. We sample the dust parameter, $\bar{\tau}_{V}$ uniformly from a normal distribution centered on zero with a standard deviation of 0.5, truncated between 0 and 4.

Since the metallicity of galaxies depends on the metal production from stars, SFR is related to metallicity. More massive galaxies have higher metallicities on average  \citep[the mass-metallicity relation; e.g][]{lequeux1979,tremonti2004}, and the scatter in this relation depends strongly on SFR. The relation between $M_{\rm gal}$--$Z_{\rm met}$--$\psi$ is known as the fundamental metallicity relation \cite[FMR; e.g.][]{mannucci2010,hunt2016}, and does not display any significant redshift evolution up to at least redshift $z>3$ \citep[e.g.][and references therein]{cresci2019}. Given these observed trends, we first assign realistic SFRs for galaxies, before assigning  metallicities from the FMR.

To assign realistic SFRs, we use the parameterization from \cite{schreiber2017}, constructed to match observations in the redshift range $0<z<7$. For SFGs, 
\begin{equation}
\begin{gathered}
\log_{10} [\psi/(M_\odot {\rm yr} ^{-1})] =\\
 \log_{10} (M_{\rm gal}/M_\odot) - 9.5 + 1.5 \log_{10} (1+z) \\
 0.3 [\max(0,\log_{10} (M_{\rm gal}/M_\odot) - 9.36 - 2.5 \log_{10} (1+z) )]^2
\end{gathered}
\end{equation}
with a log-normal scatter of 0.3 dex. 

We then propose  $Z_{\rm met}$ by drawing from the FMR presented in \cite{williams2018} \citep[based off of data from][]{hunt2016}
\begin{equation} \label{eq:Zmet}
\begin{gathered}
\log_{10}(Z_{\rm met}/Z_\odot)  + 8.7 
\approx 12 + \log_{10}({\rm O}/{\rm H})  \\
=-0.14 \log+{10}( \psi / M_{\odot} {\rm yr}^{-1} ) \\
\hspace{3cm}+ 0.37 \log_{10}(M_{\rm gal}/M_\odot) + 4.82
\end{gathered}
\end{equation}
with scatter from a Student's $t$-distribution with 3 degrees of freedom, a standard deviation of $0.3$ in $\log_{10}(Z_{\rm met}/Z_\odot)$, and truncated between  $-2.2<Z_{\rm met}<0.24$. A Student's $t$-distribution is more heavily tailed than a Gaussian, and agrees better with observational data. As discussed in \cite{williams2018}, imposing a maximum metallicity, $\log_{10}(Z_{\rm met}/Z_{\odot})=0.24$, in the synthetic galaxies reproduces the observed turnover in the mass--metallicity relation, despite the fact that Equation~\ref{eq:Zmet} is linear in mass.  We use the same FMR to assign metallicities to both SFGs and QGs, as we find that existing measurements of quiescent galaxies are consistent with the FMR used to create this synthetic catalog \citep[e.g][]{peng2015,leethochawalit2018}. 

Gas-phase metallicity is also an indirect tracer of gas flows in galaxies. Metallicity and stellar mass both correlate with star formation history, but gas flows will introduce scatter in the metallicity and stellar mass relation. Thus, the gas ionization parameter, $U_S$ (which represents the ratio of the number density of ionizing photons to the number density of hydrogen atoms) correlates with metallicity. To assign the gas ionization parameter, we use the low-redshift relation from \cite{carton2017}:
\begin{equation}
\log_{10} U_S = -0.8 \log_{10}(Z_{\rm met}/Z_{\odot})-3.58
\end{equation}
with a scatter of 0.3 dex sampled from a Student's $t$-distribution with three degrees of freedom, truncated between $-4<\log_{10}{U_S}<-1$. As in \cite{williams2018}, we assume the relation holds at all redshifts.

Given the seven FSPS parameters, we then calculate $M_{\rm UV}$, $\beta$, $\psi$ as outlined in Section~\ref{sec:calculated}. We discard galaxies that have unrealistic $\psi$, $M_{\rm UV}$ or $\beta$ values, as summarized in Table~\ref{tab:SFGs}. The resulting SFG parent catalog has $ \sim 9 \times 10^5$ galaxies.

\subsubsection{Quiescent galaxy parent catalog}

We generate the parent catalog for the QGs in a similar way to the parent catalog for the SFGs.  For $10^7$ galaxies, we assign galaxy masss, redshifts, and SFH parameters $t_{\rm start}$ and $\tau$  from uniform distributions, as summarized in Table~\ref{tab:QGs}.  We sample the dust parameter, $\bar{\tau}_{V}$ uniformly from a normal distribution centered on zero with a standard deviation of 0.5, truncated between 0 and 4.

We assume QG metallicities, $Z_{\rm met}$, follow the same FMR as SFGs, as given in Equation~\ref{eq:Zmet}, which is consistent with existing measurements of quiescent galaxies \citep[e.g][]{peng2015,leethochawalit2018}. We propose SFRs from \cite{schreiber2017} as before. For quiescent galaxies,
\begin{equation}
\log_{10} [\psi/(M_\odot {\rm yr} ^{-1})] = 0.5 \log_{10} (M/M_\odot) + \log_{10} (1+z) -6.1 \,\,\,,
\end{equation}
with a log-normal scatter of 0.45 dex. 

As before, given the seven FSPS parameters, we then calculate $\psi$ and the UVJ colors, as outlined in Section~\ref{sec:calculated}. We remove any galaxies in the QG parent catalog that do not fall in the proper space in the UVJ diagram, to avoid including any galaxies with unrealistic colors.  Specifically, we use the criteria that a galaxy can be classified as SF if it satisfies any of the following conditions
\begin{equation}
\begin{aligned}
U-V &<1.3 \\
V-J &> 1.6 \\
U-V &< 0.88(V-J) + 0.49
\end{aligned}
\end{equation}
\citep{williams2009}. 
After cuts in SFR and UVJ, the resulting parent catalog has $5 \times 10^4$ galaxies. Though SFR not set explicitly for QGs, these cuts ensure that QGs have a low SFR, differentiating them from the SFG population (see Appendix~\ref{sec:galprops}).

\subsection{Star-forming galaxies} \label{sec:SFGs}

This section outlines our procedure for assigning realistic SEDs for SFGs. To accurately capture the observed CSFRD and UVLFs, galaxies must have realistic SFRs and UV magnitudes. We also aim to produce realistic UV continuum slopes, $\beta$, the fundamental mass relation (FMR), and the relationship between metallicity and gas ionization. Overall, we propose five parameters to define the SFG SEDs, $\psi$, $M_{\rm UV}$, $\beta$, $Z_{\rm met}$, and $U_S$. Details of how these proposed parameters are generated, and how the FSPS parameters are determined for the SFGs are given in this section, and summarized in Table~\ref{tab:SFGs}.

To reproduce observed UVLFs, the synthetic galaxies need to follow the same $\bar{M}_{\rm UV}$--$M_{\rm gal}$  relation that \cite{williams2018} used to generate the high-redshift SMFs used in this work:
\begin{equation} \label{eq:MUV_W18}
\begin{aligned}
\bar{M}_{\rm UV} &= -1.66 (M_{\rm gal} - b_{M_{\rm UV}}) -20  \\
b_{M_{\rm UV}}&= 
\begin{cases}
0.12 z^2 - 0.98 z + 11.4175 &{\rm for } \,z \leq 3.75 \\ 
-0.12 z + 9.88 & {\rm for } \,  3.75<z<8 \\ 
8.92 & {\rm for } \,  z \geq 8 \\
\end{cases} 
\end{aligned}
\end{equation}
Additionally, \cite{williams2018} found that scatter observed in 3D-\emph{HST} data \citep{skelton2014} is constant in both stellar mass and redshift, with an average value of $\sigma_{\rm UV}=0.7$. Therefore, we propose $M_{\rm UV}$ values from Equation~\eqref{eq:MUV_W18} with a Gaussian scatter of $\sigma_{\rm UV}=0.7$.

We also use the relations from \cite{williams2018} to propose UV continuum slopes, $\beta$, 
\begin{equation}
\begin{aligned}
\beta &= m_\beta (M_{\rm UV} + 19.5) + b_\beta \\
b_\beta&=
\begin{cases}
 -0.09z - 1.49 & {\rm for } \,  z \leq 8\\
-0.77 & {\rm for } \,  z > 8 \\
\end{cases} \\
m_\beta&= 
\begin{cases}
-0.007z - 0.09 & {\rm for } \,  z \leq 8\\
-0.146 & {\rm for } \,  z > 8\,\,\,.
\end{cases}
\end{aligned}
\end{equation}

Given the proposed galaxy properties, we next determine which FSPS parameters to assign. To generate SEDs that match closely to the proposed parameters, we find the 10 nearest neighbors the SFG parent catalog. Table~\ref{tab:SFGs} outlines the distance metric we use to find the nearest neighbors. We average the five free FSPS parameters ($Z_{\rm met}$, $t_{\rm start}$, $\tau$, $\bar{\tau}_v$ and $U_S$)  of these neighbors, with a weight based on their distance to the proposed parameters. In cases where the assigned $t_{\rm start}$ value is greater than $t_{\rm age}$, we set  $t_{\rm start}=t_{\rm age}$.  This method allows us to ``fit" the target (proposed) parameters, even with the very large number of galaxies, something that would not be computationally realistic to achieve iteratively.

Given the large degeneracies between different galaxy properties (e.g. age, metallicity, dust, star formation history) we find that our solution depends sensitively on how realistically we populate the parent catalogs, and also how we define the distance metric. In particular, we find that to ensure the final parameters were close to the proposed parameters, it is especially important to find neighbors that are close in redshift.

\subsection{Quiescent galaxies} \label{sec:QGs}

In addition to SFGs, we include a QG population. We turn off the nebular emission model for QGs, as low redshift galaxies are known to have very low gas content. Though galaxy modeling often assumes QGs do not contain any gas or dust \citep[e.g.][]{williams2018}, recent results from \cite{gobat2018} suggest that higher redshift galaxies should have significant dust content. We therefore include dust for the QGs. We assign QG properties in a similar we assign SFG properties. Given the QG parent catalog described in Section~\ref{sec:parent}, we determine FSPS parameters  using the 10 nearest neighbors in mass and redshift. We summarize the details of this procedure in Table~\ref{tab:QGs}.

\subsection{Example SEDs}

\begin{figure}
	\includegraphics[width=\columnwidth]{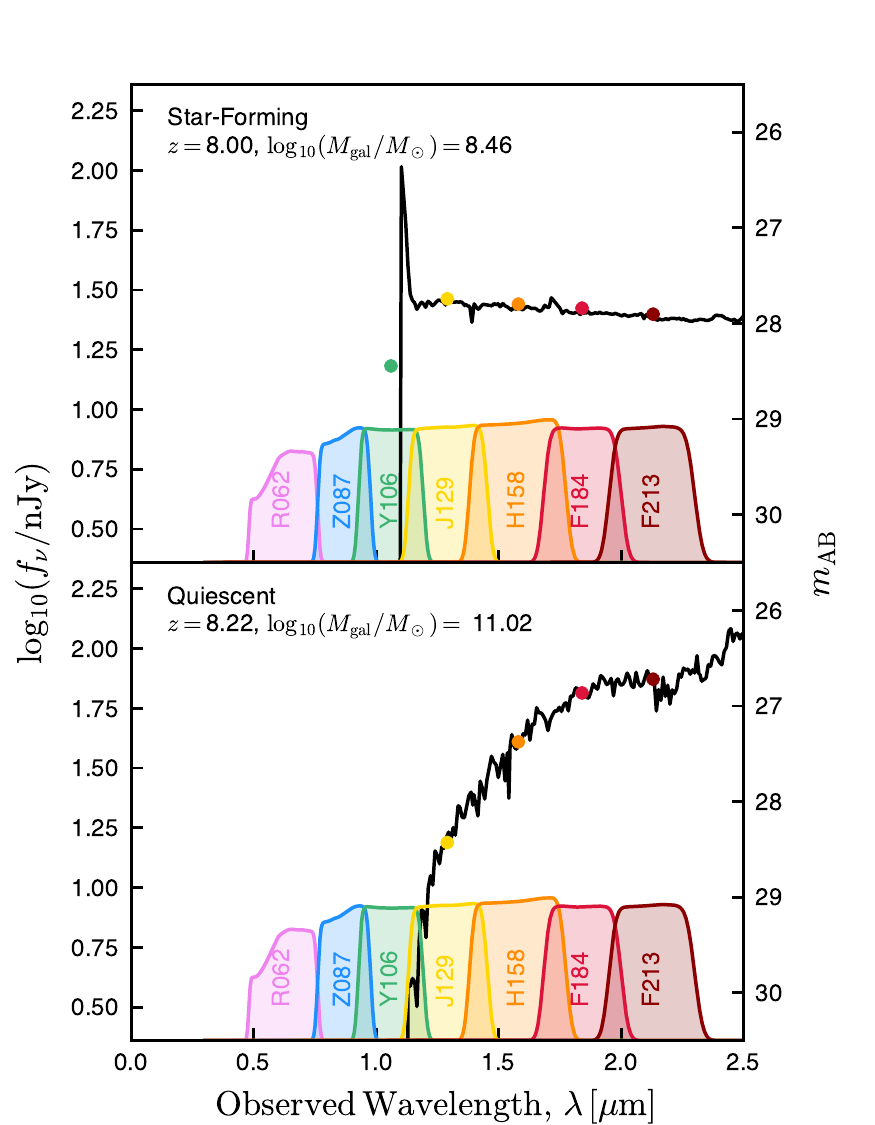}
	\caption{Example SEDs for a SFG (top) and a QG (bottom) at redshift $z \approx 8$. The Roman photometric filters (R062, Z087, Y106, J129, H158, F184,  and F213) are over-plotted. The observed magnitudes in each band are shown as colored points.  The SFG would be selected in this UDF as a LBG using \emph{Roman} photometry.}
	\label{fig:SEDs}
\end{figure}

We present the resulting SEDs for two example galaxies at redshift $z\sim8$ in Fig.~\ref{fig:SEDs}. The \emph{Roman} WFI filters include seven photometric filters: R062, Z087, Y106, J129, H158, F184,  F213\footnote{The F213 was recently added; the science justifications for this filter are outlined in \cite{stauffer2018}. The higher background in F213 will result in brighter flux limits compared to bluer Roman bands.}, a wide-field filter W146, a grism, and a prism. We plot the flux calculated in the seven photometric filters R062, Z087, Y106, J129, H158, F184, and F213. Both galaxies show a clear Lyman break in Y106.

The SFG (top panel) should be selected as a Lyman break galaxy (LBG)  in \emph{Roman} using the common Lyman-break method. This example also displays strong Ly-$\alpha$ emission. Ly-$\alpha$ emitters (LAEs) are a common way to study study the EoR, and \emph{Roman} will detect many such galaxies. Quantifying the science returns of LAEs in a \emph{Roman} UDF will be the subject of future work.

Due to the SMFs we used to construct our catalog, we model a small population of high-redshift QGs, such as the one shown in the bottom panel of Fig.~\ref{fig:SEDs}. It is unclear whether redshift $z=8$ galaxies exist, or what the mechanism of their formation would be. Due to the large area a deep survey with \emph{Roman} will cover, a \emph{Roman} UDF could help identify rare high-redshift quiescent galaxies if they exist. This will be discussed further  in Section~\ref{sec:discussQGs}.

\section{Cosmic star formation history} \label{sec:csfh}

Since one of the main goals of a \emph{Roman} UDF is to study the ionizing photon contribution from galaxies, we  ensure we reproduce observed UVLFs (Section~\ref{sec:UVprop})  and the CSFRD (Section~\ref{sec:csfrd}). Additional galaxy properties are discussed in Appendix~\ref{sec:galprops} to show the range of scientific questions our synthetic galaxy catalog can address.

\subsection{UV properties} \label{sec:UVprop}

Placing strong constraints on the faint end of the UVLF during the EoR is a main goal of a \emph{Roman} UDF. To reproduce observed UVLFs, the synthetic galaxies must follow the correct distribution of $M_{\rm UV}$, as discussed in Section~\ref{sec:UVproperties}.  On average, the UV magnitude, $M_{\rm UV}$, decreases (becomes brighter) with galaxy mass, $M_{\rm gal}$ \cite[e.g.][]{stark2009}. For galaxies with  $\log_{10}(M_{\rm gal}/M_{\odot})>10$ the relationship flattens \citep[e.g.][]{spitler2014}.  Apart from this flattening at high masses, the observed $M_{\rm UV}$--$M_{\rm gal}$ relation has a constant slope, with the normalization evolving with redshift  \cite[e.g.][]{duncan2014,stefanon2017}. 
Similarly, the UV continuum slope, $\beta$, correlates with $M_{\rm UV}$. Though the exact relation between $\beta$ and $M_{\rm UV}$ is still being studied, $\beta$ appears bluer (decreases) with increasing (fainter) $M_{\rm UV}$ and increasing redshift \cite[e.g.][]{bouwens2014}.

\begin{figure}
	\includegraphics[width=\columnwidth]{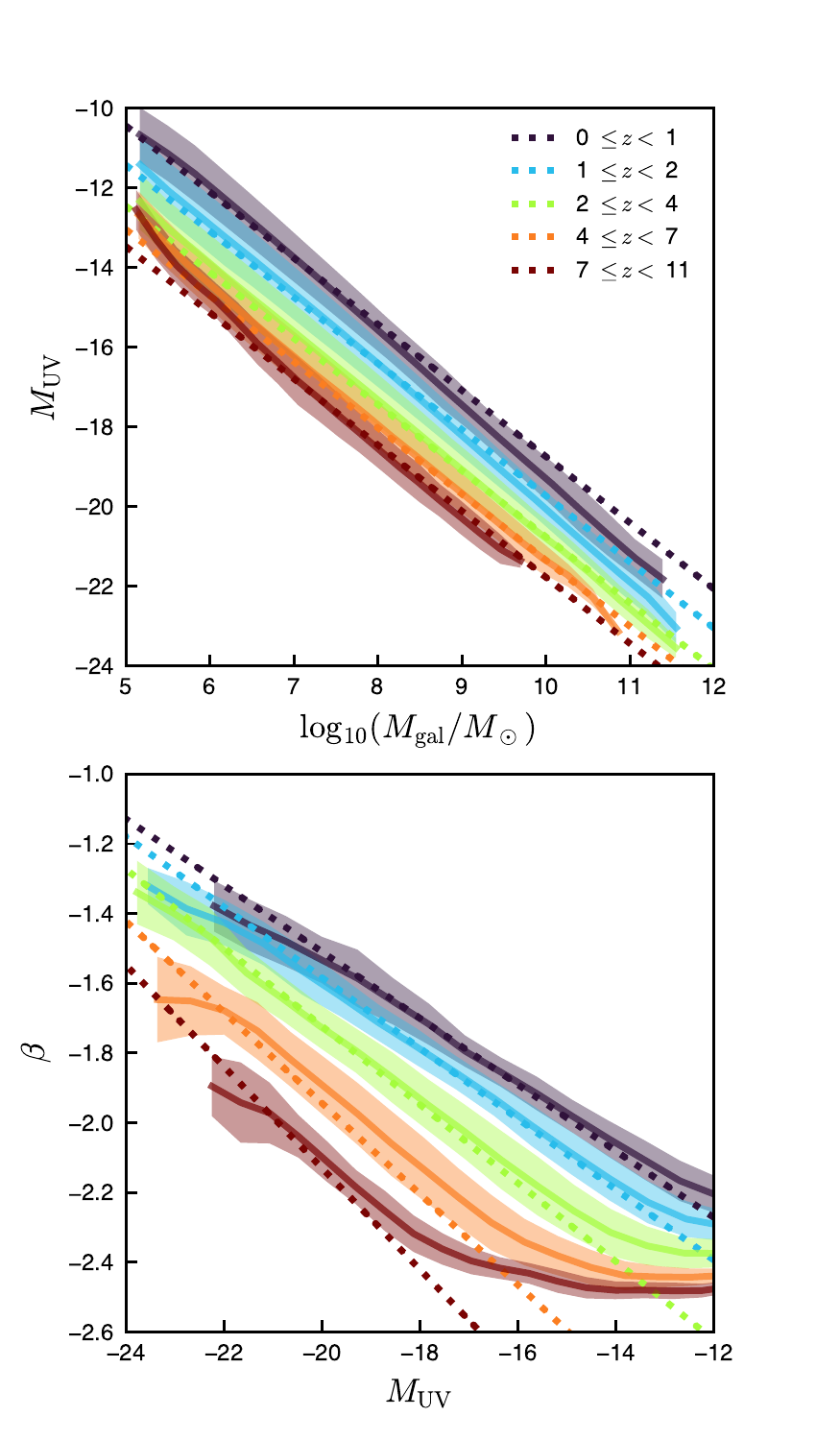}
	\caption{Average $M_{\rm UV}$--mass relation (top) and $\beta$--$M_{\rm UV}$ relation (bottom) for SFGs in the synthetic catalog (solid lines) and the standard deviation in the bin (shaded region).  We compare the synthetic catalog to the relations from \cite{williams2018}, calculated as the volume-weighted average over the redshift bin (dotted lines). The synthetic galaxy catalogs are constructed to match the underlying $M_{\rm UV}$--mass relations, to reproduce observed UVLFs. The synthetic galaxy catalog matches the underlying relations closely, except for a flattening in the $\beta$--$M_{\rm UV}$ relation for faint galaxies. This flattening is consistent with current constraints, as described in Section~\ref{sec:UVprop}.}
	\label{fig:MUV}
\end{figure}

We calculate the UV properties, $M_{\rm UV}$ and $\beta$, of the synthetic SFGs as described in Section~\ref{sec:UVproperties}, and presented in Fig.~\ref{fig:MUV}. We reproduce trends for both $M_{\rm UV}$ and $\beta$ imposed from \cite{williams2018}. The $M_{\rm UV}$-mass relation enables us to reproduce observed UVLFs, as discussed below. We also closely reproduce the $\beta$--$M_{\rm UV}$ relation for $\beta>-2.5$. Below $\beta<-2.5$, our galaxy catalog agrees with studies that indicate that the $\beta$--$M_{\rm UV}$ relationship flattens for $M_{\rm UV} \gtrsim -19$ galaxies \citep[e.g.][]{bouwens2014}.

\begin{figure}
	\includegraphics[width=\columnwidth]{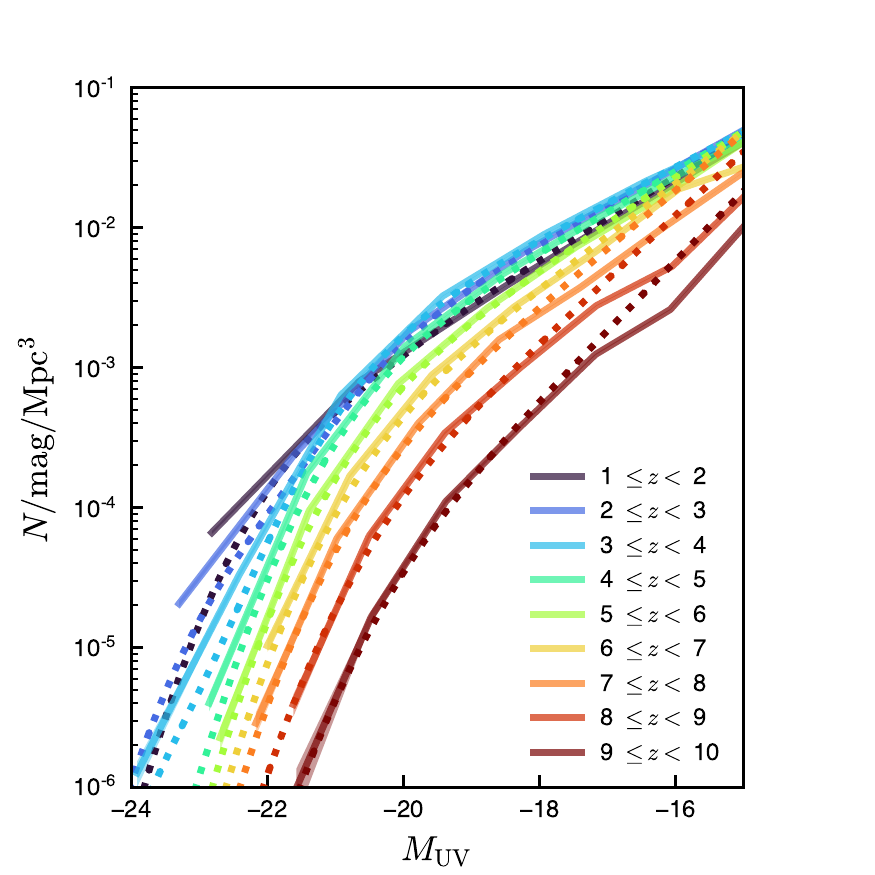}
	\caption{UV luminosity function of the synthetic galaxy catalog. The dotted lines show the expected UVLF from convolving the \citet{williams2018} SMF with the  $M_{\rm UV}$--$M_{\rm gal}$ relationship (Equation~\ref{eq:UVLF}), and then averaged over the redshift bin. Overall, the data matches the underlying relations very well, and agrees with current observations (see Fig.~\ref{fig:LF_catalog}).}
	\label{fig:LF}
\end{figure}

We show the UVLFs of our synthetic SFGs in Fig.~\ref{fig:LF}, compared to the UVLFs that are expected from convolving the SMF with a normal distribution, centred on $M_{\rm UV}$ (Equation~\ref{eq:UVLF}). The galaxy catalog agrees very well with the imposed relation, and is consistent with current observations, as shown in Section~\ref{sec:projected}. Our catalog has slightly less galaxies at faint magnitudes ($M_{\rm UV}>-17$) than expected due to the fact that we do not perfectly recreate the $M_{\rm UV}$--$M_{\rm gal}$ relationship. These magnitudes are fainter than the detection limits of a $30\,m_{\rm AB}$ survey, and our synthetic catalog UVLF is consistent with current observations, as discussed in Section~\ref{sec:projected}.

\subsection{Cosmic star formation rate density} \label{sec:csfrd}

\begin{figure} 
	\includegraphics[width=\columnwidth]{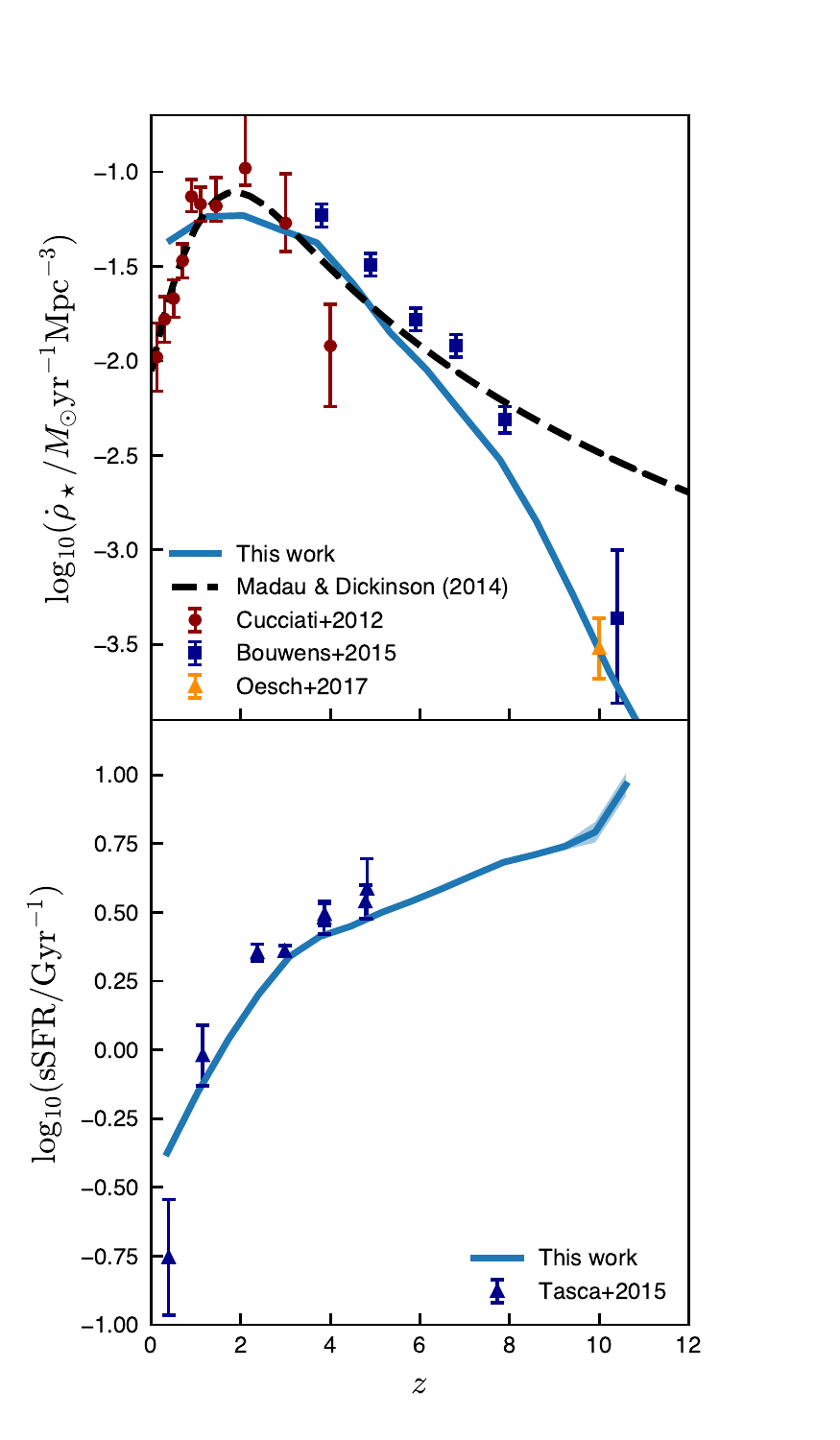}
	\caption{Evolution of star formation in the universe. Top: evolution of the CSFRD of synthetic galaxies with a luminosity limit of $0.03 \, L_*$ (dark blue). Error bars are smaller than the width of the line. The dashed black line is the CSFR density of the universe compiled by \cite{madau2014}, and colored points are measurements from the literature \citep{cucciati2012,bouwens2015,oesch2018}. The data are converted from a \cite{salpeter1955} to a \cite{chabrier2003} IMF by dividing the CSFRD by a factor of 1.7. Bottom: the redshift evolution of the median sSFR  of $8.8 < \log_{10} (M_{\rm gal}/M_{\odot}) <10$  synthetic galaxies (solid line) compared to measurements from the literature \citep[colored points;][]{tasca2015}. Overall, the cosmic star formation history of the synthetic galaxy catalogs is in close agreement with observational data.}
	\label{fig:SFR_vs_z}
\end{figure}

As discussed in  Section~\ref{sec:SFH}, the CSFRD describes the ionizing photon contribution of galaxies as a function of cosmic time \citep[for a review, see e.g.][]{madau2014}. We display the evolution of the CSFRD in the top panel of Fig.~\ref{fig:SFR_vs_z}. We calculate the CSFRD by taking the sum of galaxy SFRs, $\psi$ (averaged over the last 100 Myr), in each bin divided by the comoving volume of the bin. As in \cite{williams2018}, to compare to \cite{madau2014}, we imposed a luminosity limit of $0.03 \, L_*$.\footnote{The equivalent limit in $M_{\rm UV} $ is $M_{\rm UV} \sim -14.5$ at $z<1$ \citep{cucciati2012},  $M_{\rm UV} \sim -15.5$ at $1<z<2$ \citep{cucciati2012},  $M_{\rm UV} \sim -16.89$  at  $2<z<2.7$ \citep{reddy2009}, and $M_{\rm UV} \sim -17$  at $z>2.7$ \citep{bouwens2015,bouwens2016,finkelstein2015,oesch2018}.}

Similar to \cite{williams2018}, we find that the CSFRD agrees roughly with the model presented in \cite{madau2014}, but more closely to data \citep[e.g.][]{cucciati2012,oesch2018}. Most notably, our synthetic galaxies agree with high-redshift ($z>8$) measurements of the CSFRD. As in \cite{williams2018}, our synthetic galaxies have a  slight deficit at redshifts $1<z<3$, which reflect the underlying SMFs. As discussed in \cite{williams2018},  the SMFs we used do not include the dusty SFGs currently missed in the UV-selected samples.

We show the evolution of the specific star formation rate (sSFR) in the bottom panel of Fig.~\ref{fig:SFR_vs_z}, for galaxies with $8.8 < \log_{10} (M_{\rm gal}/M_{\odot}) <10$. We calculate the median sSFR by dividing the SFR averaged over the last 100 Myr by the galaxy mass. For comparison, we plot data from \cite{tasca2015}. The synthetic galaxies and observations agree very well. The sSFR of galaxies at redshifts $z>4$ is a matter of active research, but we find an increase in sSFR, consistent with current studies \citep{stark2016}. Given that our galaxy catalog is consistent with observations for both galaxy clustering and cosmic star formation histories, we now turn our attention to making preliminary predictions for the science returns of a \emph{Roman} UDF.

\section{Predictions for a Roman ultra-deep field survey} \label{sec:predicts}

In this section, we demonstrate the capabilities of a $1\,\deg^2$  \emph{Roman} UDF to study the earliest galaxies at the EoR and beyond. We begin by discussing our method for determining whether a galaxy is selected (Section~\ref{sec:detect}), and the expected galaxy number counts (Section~\ref{sec:numcounts}), before presenting our projected constraints (Section~\ref{sec:projected}). Throughout this section we compare our predictions for a \emph{Roman} UDF  to the best current constraints in the literature. However, by the time \emph{Roman} launches \emph{JWST} will have provided an extraordinary amount of data, greatly enhancing our understanding of the universe. We discuss synergies between a \emph{Roman} UDF and \emph{JWST} in Section~\ref{sec:jwst}.

\subsection{Selection criteria} \label{sec:detect}

Understanding of galaxy formation increased immensely when techniques to select high redshift ($z>5$) galaxies were developed  \citep[for review see, e.g.][]{dunlop2013}. The most commonly used method is the Lyman-break technique, which selects LBGs based on a step in the blue UV continuum emission at $\lambda_{\rm rest}=1216$\,\AA. To determine whether  a UDF with \emph{Roman} would select each galaxy  in the synthetic catalog, we use a magnitude limit based on the Lyman-break, in the same manner as  \cite{williams2018}.  Specifically, for each filter, we assign a non-overlapping wavelength range (the dividing point was taken to be halfway between the centres of two adjacent bands). Each galaxy is assigned a ``dropout" filter, according to the wavelength that corresponds to its Lyman-break, $\lambda=1216(1+z)$\AA. If a galaxy is brighter than the magnitude limit in the band immediately redder to its dropout filter, we consider it to be detectable. We use the magnitude limit  $m_{\rm AB}=30$ for a  $5\sigma$ detection.

\begin{figure}
	\includegraphics[width=\columnwidth]{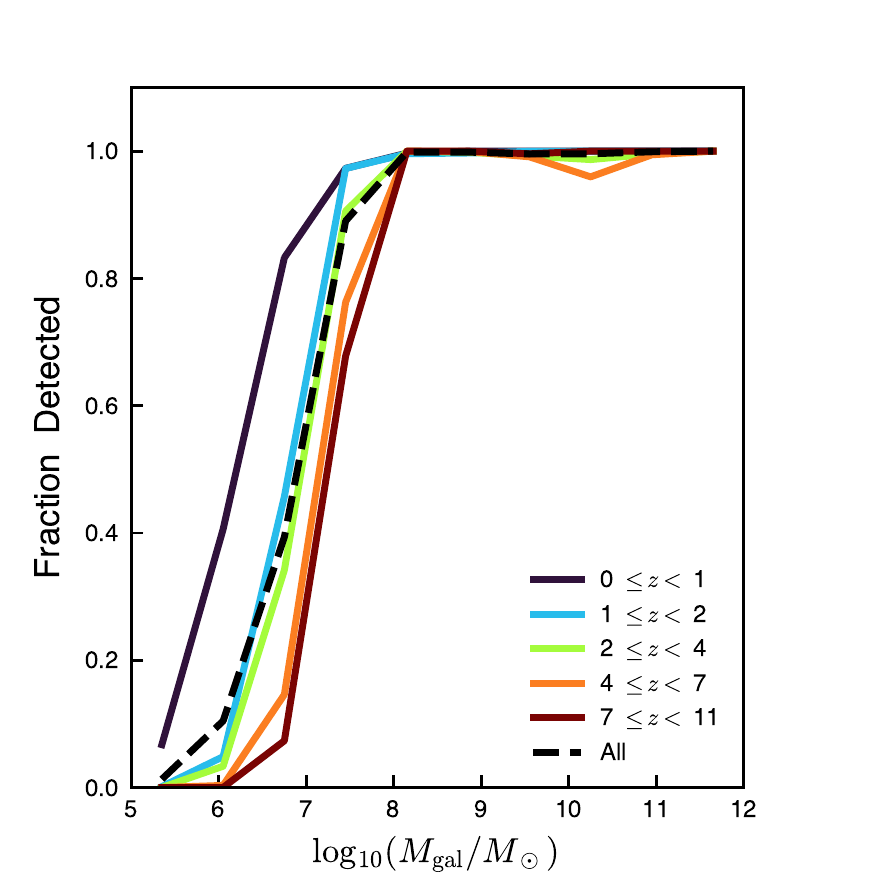}
	\caption{Fraction of detectable galaxies ($5 \sigma$) in the Roman synthetic catalog. Nearly all galaxies with masses $M_{\rm gal} > 10^8 M_{\odot}$  (corresponding to $M_{\rm UV} < -18$) are detectable out to redshifts $z\approx10$.}
	\label{fig:frac_detect}
\end{figure}

We show the fraction of detectable galaxies  in Fig.~\ref{fig:frac_detect}. The mass at which $\sim 50$\% of galaxies ranges from $\sim10^{6}$ at $z \approx 1$ to $10^7$ at $z=10$. We find that nearly 100\% of galaxies are detectable above masses of $M_{\rm gal} = 10^8 M_\odot$, which indicates that there should be very tight constraints on galaxy properties (such as galaxy clustering and UVLFs) above this mass. Given this, we can measure the clustering of faint galaxies ($M_{\rm UV} \lesssim -18$) at the EoR.  With completeness corrections, we can measure quantities such as the UVLFs for galaxies brighter than $M_{\rm UV} \approx -17$ out to redshifts $z\approx 10$. Section~\ref{sec:projected} will explore these potential constraints in more detail.

In the following sections, we assume that any detectable galaxy is selected. This approach assumes that there is no scatter in the photometric redshifts, and no galaxies are obscured by forefront galaxies. Future work will examine \emph{Roman} UDF constraints in more detail, taking into account these effects. The catalogs and images presented in this work will be instrumental to explore these sources of systematic errors.

\subsection{Number counts}\label{sec:numcounts}

\begin{figure}
	\includegraphics[width=\columnwidth]{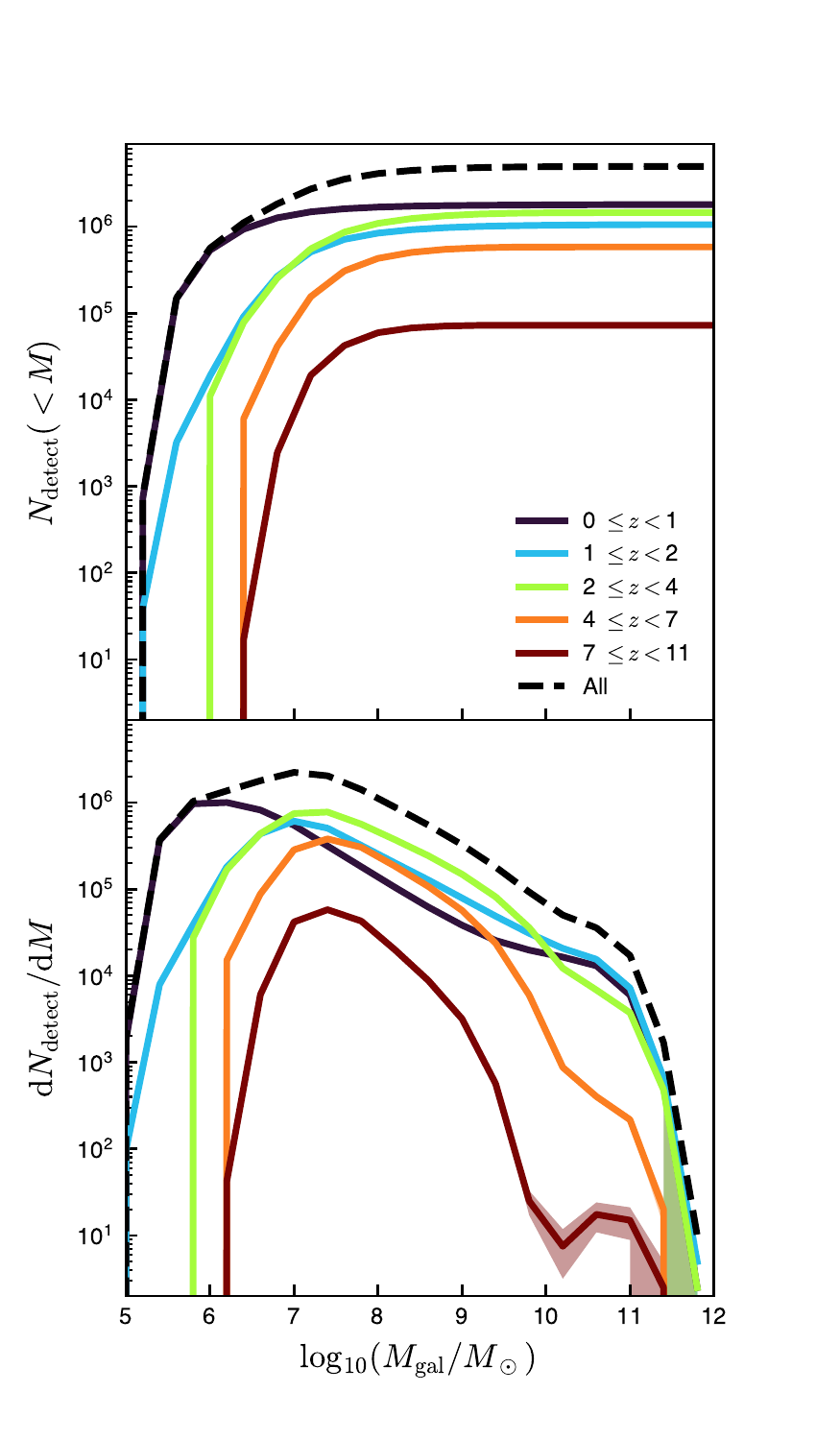}
	\caption{The cumulative (top) and differential (bottom) number counts of the detectable galaxies ($5 \sigma$) in our \emph{Roman} UDF synthetic catalog. A \emph{Roman} UDF will contain over a million galaxies, with tens of thousands in the EoR.}
	\label{fig:Ndetect}
\end{figure}

Here we provide number counts of the detectable galaxies ($5\sigma$), to understand the science returns of a \emph{Roman} UDF and measure the completeness of the survey. Fig.~\ref{fig:Ndetect} shows both the cumulative number counts (top), and the differential number counts (bottom). We find that a $1 \deg^2$ \emph{Roman} UDF could detect more than $10^4$ galaxies at redshifts $z>7$. For comparison, \emph{HST} has detected on the order of $10^3$ galaxies at redshifts $z \sim 4$--$10$ \citep{koekemoer2013,lotz2017}, and Cycle 1 \emph{JWST} programs are expected to detect $\sim 5000$ galaxies at redshifts $z>6$ \citep[][]{williams2018}.

\begin{figure}
	\includegraphics[width=\columnwidth]{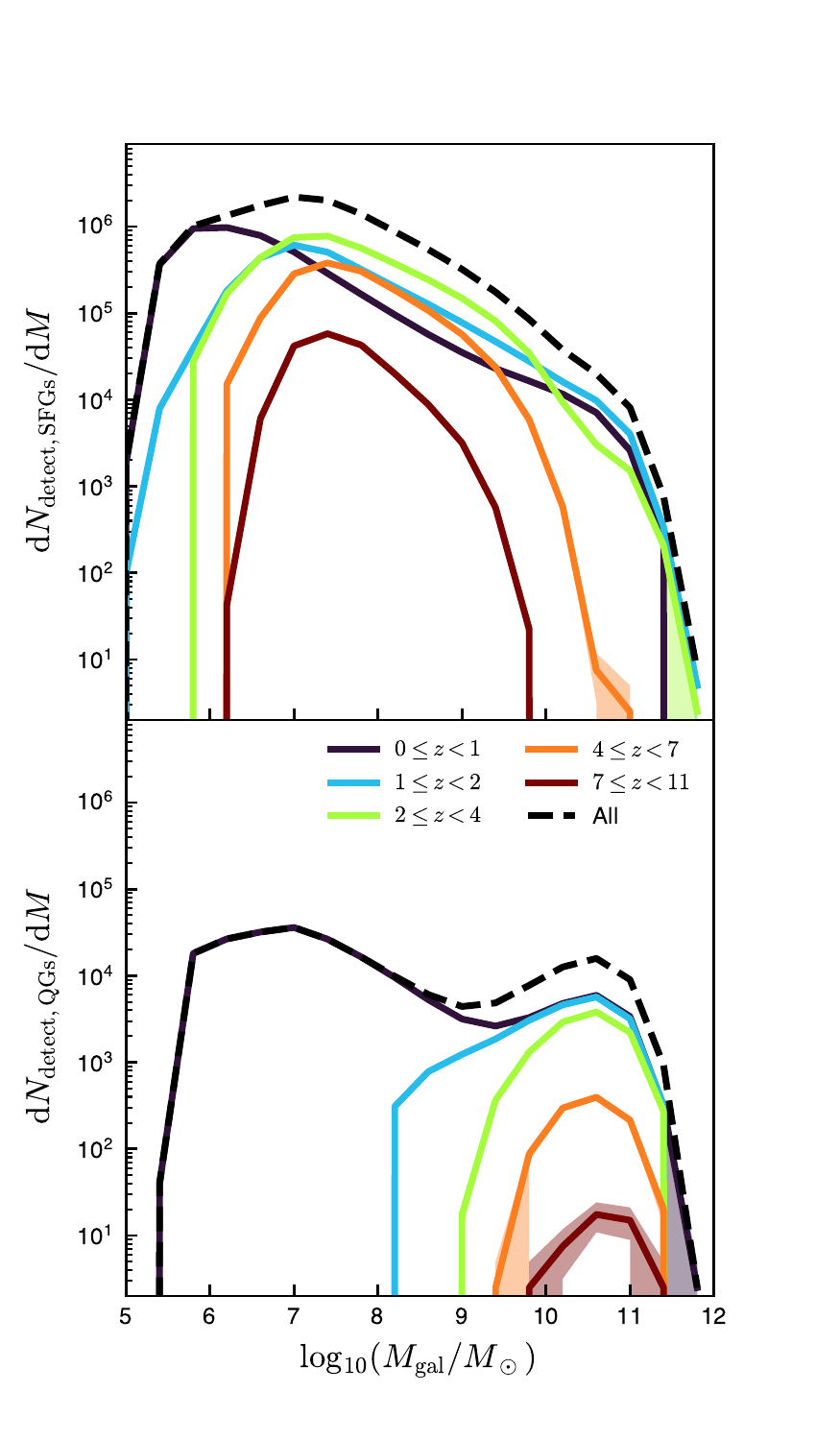}
	\caption{Number counts of detectable galaxies ($5 \sigma$) for SFGs (top) and QGs (bottom) in our \emph{Roman} UDF synthetic catalog. A \emph{Roman} UDF will detect QGs out to very high redshifts, possibly even detecting a few at the EoR.}
	\label{fig:NdetectSF}
\end{figure}

Though a UDF catalog will consist largely of SFGs, there will also be a number of QGs, as demonstrated in Fig.~\ref{fig:NdetectSF}. These predictions indicate that a $1\, \deg^2$ will contain a handful of detectable QGs above redshift $z=7$ if they exist. In combination with e.g. \emph{JWST} spectroscopy, this could identify $z>7$ QGs. Given that we currently have only detected QGs out to redshifts of $z \sim 4$--$5$ \citep{merlin2019,valentino2020}, this will greatly enhance our understanding of the QG population. Determining the number and mass distribution of high-redshift QGs is fundamental to understanding galaxy evolution, and for testing different models for the emergence of QGs \citep[see, e.g.][for a recent review on quenching]{cortese2021}.

Due to the nature of these kinds of predictions, the underlying galaxy properties in the catalog are extrapolated beyond our current knowledge (in particular, the number of low-mass objects is dependent on extrapolations of the SMFs). Therefore, the number counts presented in this section represent a reasonable estimate of what a UDF might see, but there is much to be learned in the future. Regardless, the large number of galaxies that \emph{Roman} will detect will provide the best census to date of faint high-redshift galaxies, including those thought responsible for reionizing the universe.

\subsection{The science returns of a Roman UDF} \label{sec:projected}

Some of the quantities that a \emph{Roman} UDF will aim to constrain include the SMF, the UV luminosity function,  galaxy clustering and the galaxy--halo connection at high redshifts. In this section we provide preliminary projected constraints for a $1 \deg^2$ \emph{Roman} UDF, based on the detectable galaxies identified in Section~\ref{sec:detect}. We have not included uncertainties from cosmic variance or from errors on photometric redshifts in our predictions, as  that  is beyond the scope this paper. However, we note that our synthetic catalog provides the basis for rigorous studies of systematic errors in a \emph{Roman} UDF, and will be the focus of future work. 

\subsubsection{Stellar mass functions} \label{sec:SMFpredict}

\begin{figure}
	\includegraphics[width=\columnwidth]{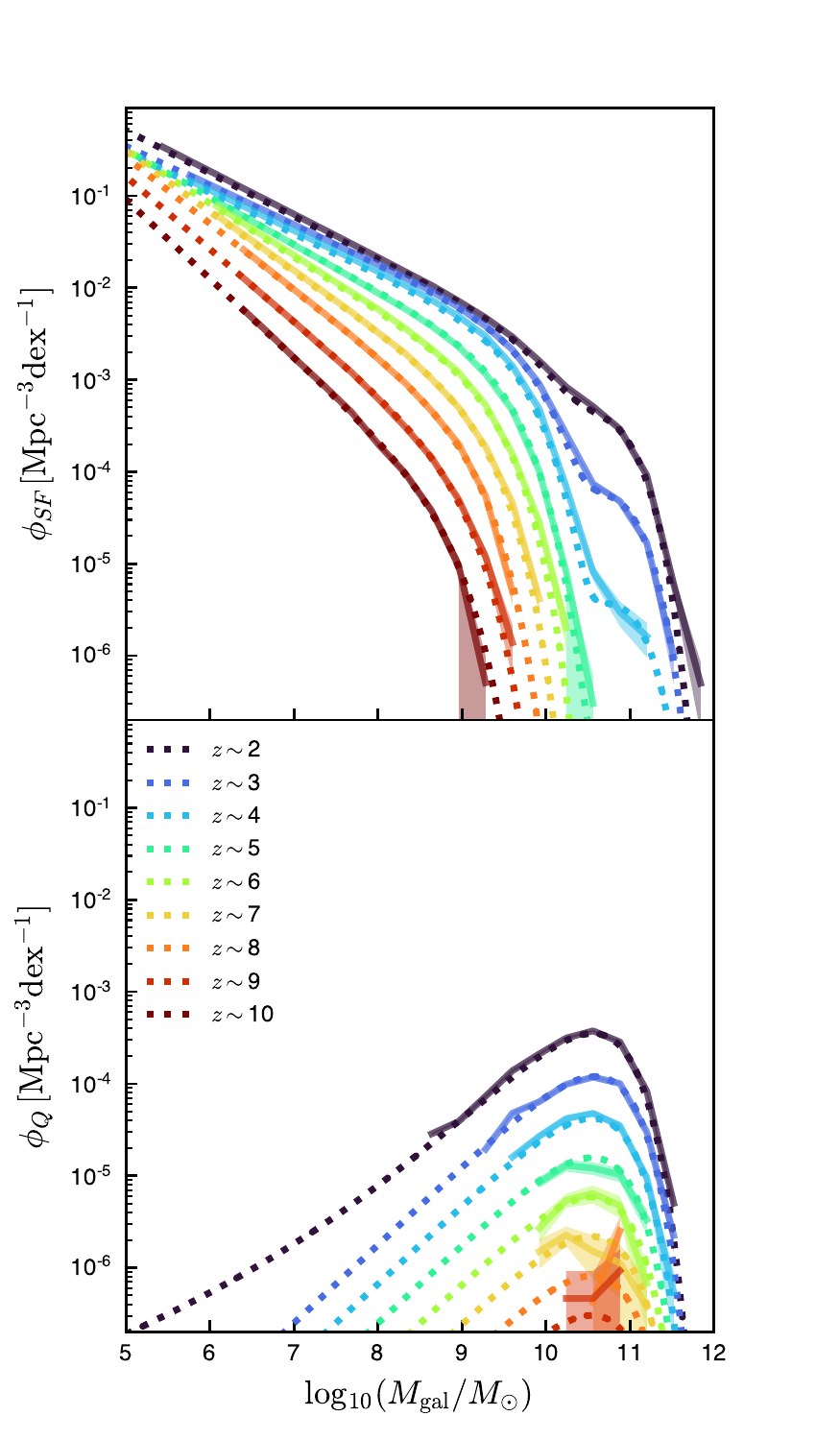}
	\caption{SMF of the detectable galaxies ($5\sigma$) in our synthetic galaxy catalog for SFGs 
(top) and QGs (bottom), with completeness corrections. The synthetic galaxies agree with the \cite{williams2018} SMFs (dotted lines) by construction. A \emph{Roman} UDF can potentially constrain the SMF beyond redshift $z=10$ ($M_{\rm gal}>10^{6.5}\, M_{\odot}$) for SFGs, and to redshift $z\approx7$ ($M_{\rm gal}>10^{9.5}\, M_{\odot}$) for QGs.}
	\label{fig:SMF_catalog}
\end{figure}

The evolution of the SMF of galaxies over cosmic time provides information on how galaxy populations have evolved, the star-formation history of the universe, and the galaxy--halo connection. Fig.~\ref{fig:SMF_catalog} shows the SMF for the detectable galaxies in the synthetic catalog, with the underlying SMFs shown as dotted lines for both SFGs (top) and QGs (bottom). We perform a completeness correction by dividing the number of detectable galaxies by the fraction of detectable galaxies at the corresponding mass and redshift. As shown in Fig.~\ref{fig:AbundanceMatching}, our synthetic catalog matches the SMF nearly exactly by construction.

For our catalog, the SMF for the detectable SFGs is measurable to masses of $M_{\rm gal}>10^{6.5}\, M_{\odot}$ for all redshifts.  For the detectable QGs, the SMF is measurable to redshifts $z\sim 7$ (masses $M_{\rm gal} > 10^{9.5}\, M_{\odot}$). In comparison, observations have only recently placed \emph{any} constraints on the SFG SMF at redshift $z=8$--$10$. Further, these constraints only exist for galaxies with $M_{\rm gal}<10^{8}\, M_{\odot}$, and have very large uncertainties  \citep[about 100\%;][]{stefanon2021}. Measurements of QG SMFs only extend to redshifts $z \approx 4$ \citep[e.g.][]{girelli2019}. 

The exact constraints on a SMF from \emph{Roman} will depend on stellar mass measurements and the accuracy of completeness corrections. This work provides a first calculation of the completeness of a \emph{Roman} UDF (Section~\ref{sec:detect}), and highly realistic simulated data to enable future rigorous calculations. In practice, stellar mass estimates of high redshift galaxies will require photometry redshifts beyond the reddest \emph{Roman} filters. However, \emph{JWST} will provide tight constraints on the  $M_{\rm UV}$--$M_{\rm gal}$ relation which can be used to estimate  stellar masses of galaxies in a \emph{Roman} UDF.

\subsubsection{UV luminosity functions}

To clearly answer whether there were enough galaxies during the EoR to reionize the universe, we need strong constraints on the faint end of the UVLF during the EoR.  \cite{bouwens2021} provide the most recent, comprehensive constraints on the UVLF, and are able to constrain the faint end ($M_{\rm UV}<-17$) out to redshift $z\sim10$. At redshift $z\sim 10$, only 8 sources are used to constrain the UVLF, so the uncertainties are very large ($\sim$50--100\%). In comparison, a \emph{Roman} UDF could detect $\sim10^3$ sources at $z\sim10$, and thus greatly reduce the uncertainties in the UVLF.

\begin{figure*}
	\includegraphics{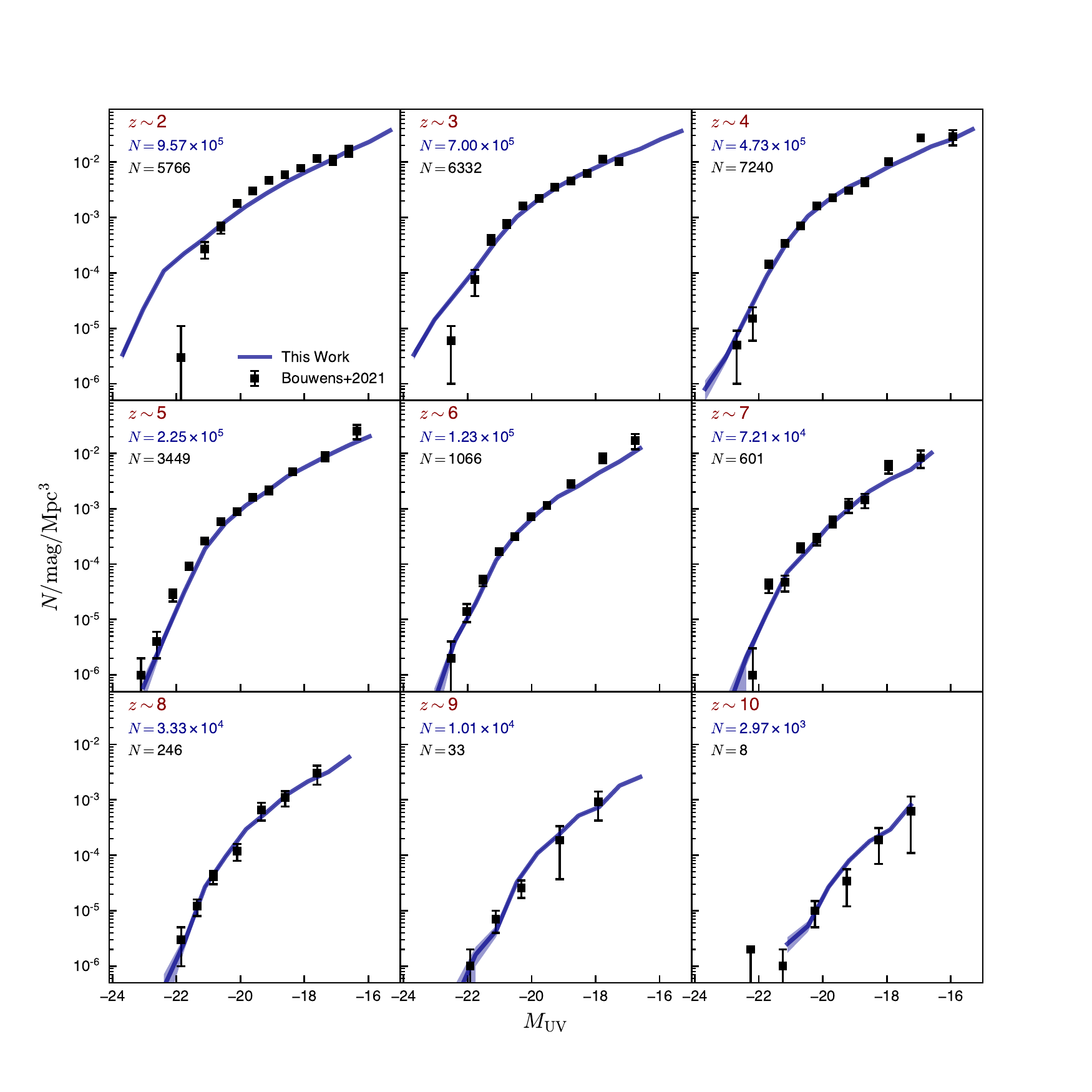}
	\caption{The UVLF of the detectable galaxies in our synthetic galaxy catalog (blue lines) and the 1-$\sigma$ Poisson noise (shaded region). We show the recent \cite{bouwens2021} data for comparison (black points). We also indicate the number of sources, $N$, in each panel used to calculate the UVLF. A \emph{Roman} UDF will provide remarkably tight constraints on the faint end of the UVLF at high redshifts.}
	\label{fig:LF_catalog}
\end{figure*}

We show the UVLF of the detectable synthetic galaxies in Fig.~\ref{fig:LF_catalog}, compared to the data from \cite{bouwens2021}. As in the previous section, we perform a completeness correction by dividing the number of selected galaxies by the fraction of selected galaxies at the corresponding mass and redshift. We predict a \emph{Roman} UDF will be able to constrain the UVLF beyond redshift $z\approx10$, with galaxies $M_{\rm UV}<-17$. Due to the vast number of galaxies in a \emph{Roman} UDF, it could constrain the faint-end UVLF to within $\sim 1$ percent  to redshifts $z\gtrsim10$, far tighter than existing limits.

We note that the bright end of the synthetic galaxy catalog UVLF is higher than the \cite{bouwens2021} data at low redshifts. As shown in Fig.~\ref{fig:LF}, our galaxy catalog agrees with the underlying UVLF model we used at the bright end, and therefore this difference can be attributed to the SMFs used in this work. This excess of low-redshift ($z<4$), bright galaxies is consistent with the possibility of inefficient mass quenching, low dust obscuration, or hidden AGN activity, as suggested by \cite{harikane2021}.

\subsubsection{Galaxy clustering}

\emph{Roman}'s wide, contiguous field of view will allow measurements of the 2PCF of faint galaxies at the time of reionization and thus their underlying dark matter halo masses. To date, halo masses have only been strongly constrained to redshifts $z\sim 6$, and are limited to bright ($M_{\rm UV} \sim -20 $) galaxies at redshifts $z=4$--$6$ \citep[e.g.][]{harikane2018a}. JWST will potentially provide the first measurement of the clustering of high-redshift galaxies \citep{endsley2020}, but due to its smaller field of view, it will likely suffer from cosmic variance, and will not have the same capability as \emph{Roman} to study clustering in different environments.

\begin{figure*}
	\includegraphics[]{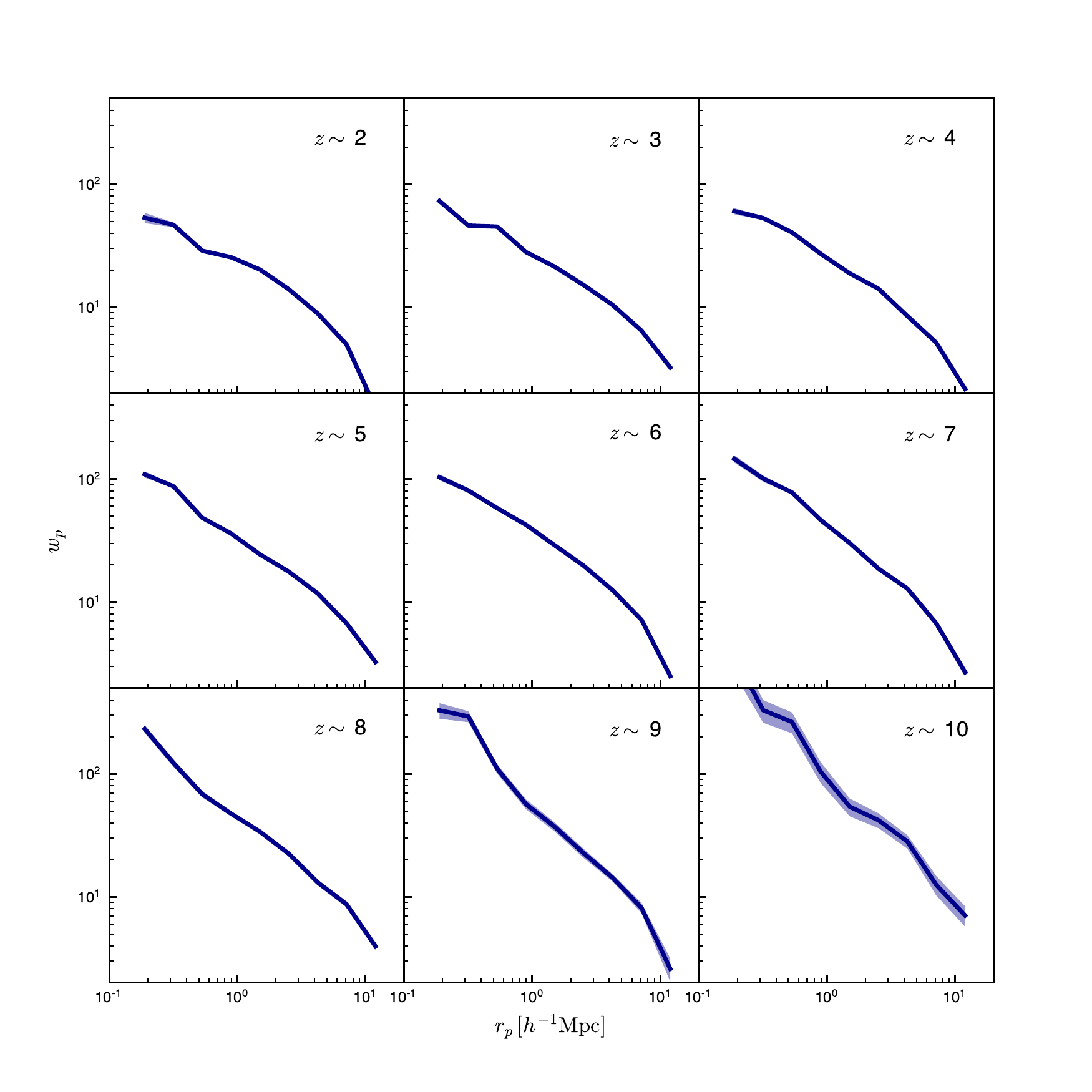}
	\caption{The 2PCF of the  $M_{\rm UV}<-18$ detectable galaxies in our synthetic galaxy catalog (blue lines), with the uncertainty calculated from bootstrapping the data. (shaded region).  The 2PCF of the detectable galaxies is measured to within  1\% for galaxies $z \lesssim 7$, and to within 10\% at  $z\approx 10$.}
	\label{fig:Clustering_catalog}
\end{figure*}

We measure the 2PCF of the detectable galaxies in the $1\deg^2$ survey, as outlined in  Section~\ref{sec:gallightcone}. We included all galaxies brighter than $M_{\rm UV}<-18$. Fig.~\ref{fig:Clustering_catalog} shows the measured 2PCF. 
The 2PCF for these faint galaxies is measured to very high accuracy (within 1\%) out to redshifts $z\approx 7$, with constraints out to $z\approx 10$ (to within 10\%).

\subsubsection{Galaxy--halo connection}

In addition to constraining the number and spatial distribution of faint, high-redshift galaxies, \emph{Roman} will also provide constraints on the galaxy--halo connection. Galaxy clustering can be used to infer dark matter halo mass, as the underlying dark matter will dictate the gravitational field \citep{mo1996}. Together with galaxy mass measurements, galaxy clustering gives a direct measure of the SHMR (which summarizes the connection between galaxy masses and their host dark matter halos). 

As discussed above, halo masses have only been measured directly for redshifts $z\lesssim6$ \cite[e.g.][]{harikane2018a}, while at high redshifts, SHMR constraints have been achieved with abundance matching between observed stellar masses and dark matter only simulations \cite[e.g.][]{stefanon2021}. There is a potential disagreement between these two techniques---the SHMR measured from abundance matching is typically 3--4 times higher than that measured from clustering \citep{stefanon2021}.  The difference between SHMRs derived from clustering and AM indicates that direct halo mass measurements are needed to understand the galaxy--halo connection at high redshifts. A \emph{Roman} UDF will provide extraordinary data, which may elucidate the origin of this discrepancy.

\begin{figure*}
	\includegraphics[]{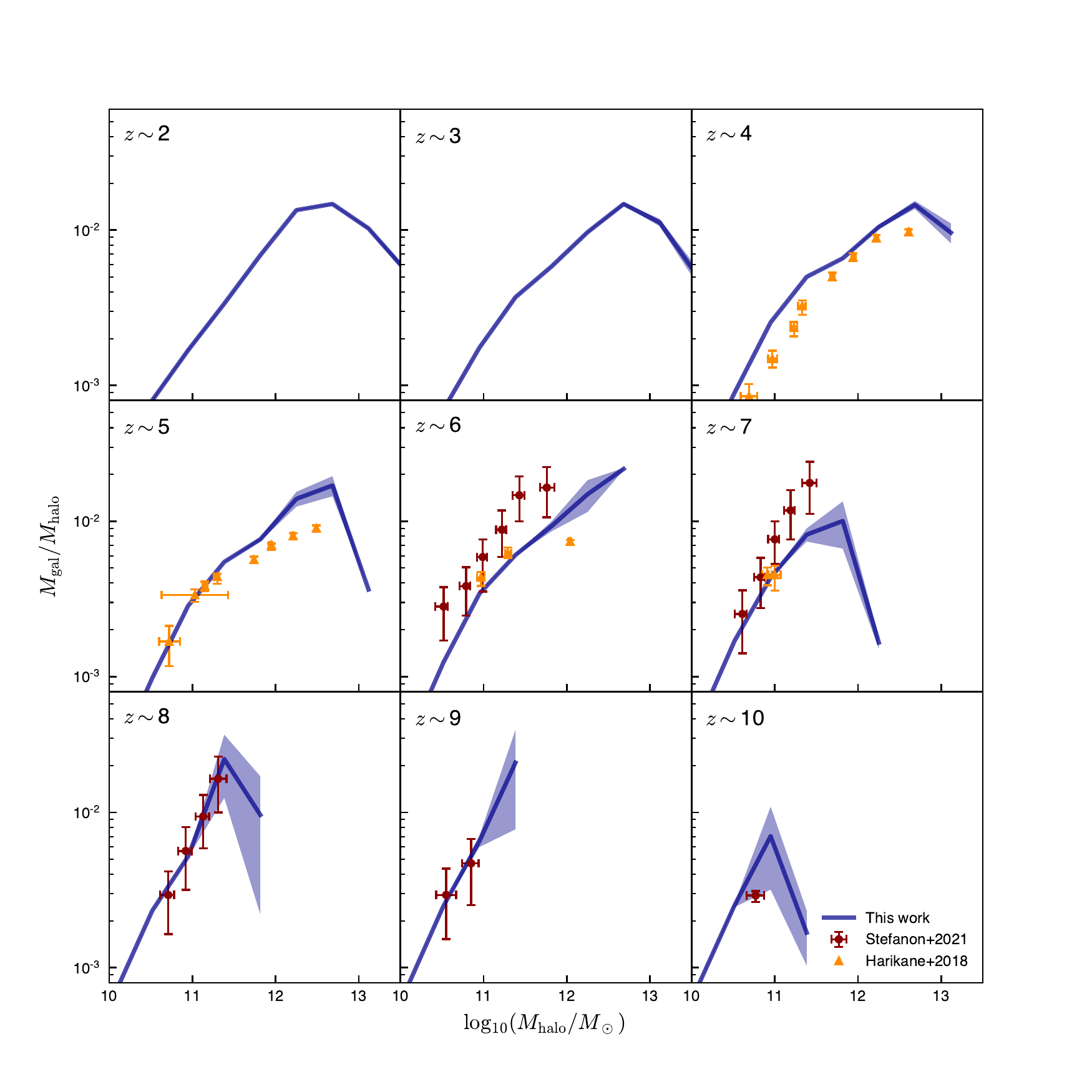}
	\caption{SHMR  of the detectable galaxies in our synthetic galaxy catalog (blue solid line), and the 1$\sigma$ spread (shaded blue region). We compare to data from \cite{harikane2018a} (orange points; error bars show error in the mean) and \cite{stefanon2021} (red points; error bars denote the scatter). The \cite{stefanon2021} data has been scaled by a factor of $1/1.7$ to convert to a \cite{chabrier2003} IMF. \cite{stefanon2021} derived halo masses from abundance matching, while \cite{harikane2018a} derived halo masses from galaxy clustering data.}
	\label{fig:SHMR}
\end{figure*}

We show the SHMR for our synthetic galaxy catalog in Fig.~\ref{fig:SHMR}, and recent constraints from \cite{harikane2018a} and \cite{stefanon2021}. The \cite{stefanon2021} data were scaled down by a factor of 1.7 to convert from a \cite{salpeter1955} IMF to a \cite{chabrier2003} IMF.  We use all the $5\sigma$ Lyman-break selected galaxies in the catalog. Our synthetic catalog clustering measurements agree well with the \cite{stefanon2021} data (within $\sim 1 \sigma$), and are slightly higher than the  \cite{harikane2018a} data.

In this work we do not attempt to predict how well a \emph{Roman} UDF will be able to constrain the SHMR. As discussed in Section~\ref{sec:SMFpredict}, stellar mass measurements for high-redshift galaxies likely require photometry at wavelengths longer than what \emph{Roman} will provide, but may be possible to estimate with scaling relations. Since our  synthetic galaxy catalogs contain information regarding host dark matter halo properties, this work can form the basis of future analyses of the galaxy--halo connection in UDFs.

\subsection{Synthetic images} \label{sec:images}

In addition to the galaxy catalog, we present  synthetic images of a \emph{Roman} UDF. These images are intended for developing analysis tools, and studying systematics (see Section~\ref{sec:discuss_img}). Full resolution versions of the RGB images presented in this section in our data release (see Appendix~\ref{sec:datarel} for details). 

\begin{figure*}
	\includegraphics[]{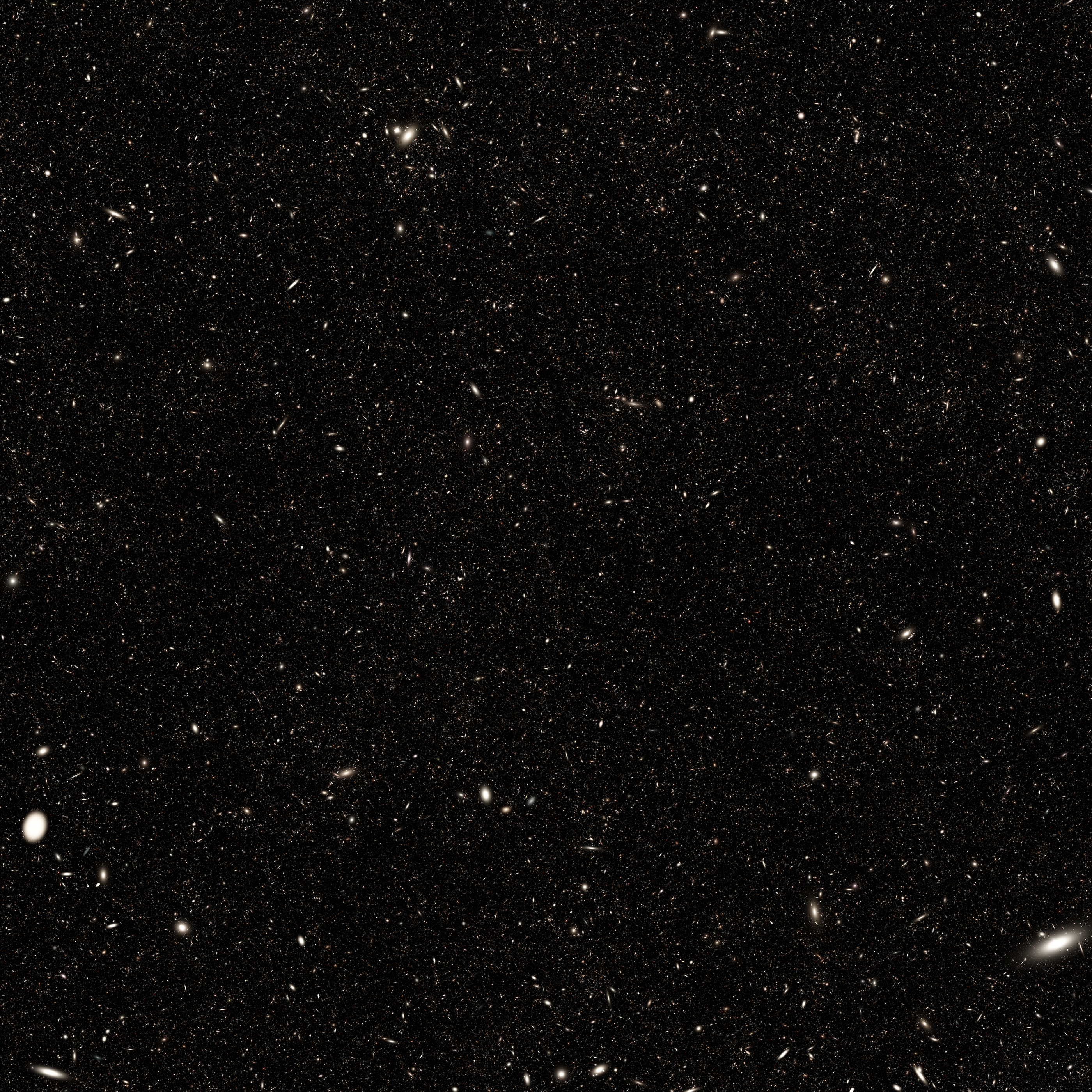}
	\caption{Noise-free composite simulated image of  the full $1\deg^2$ galaxy catalog using the Z087 (blue), Y106 (green), and H158 (red) filters. The image was created with the \emph{Roman} module in \textsc{GalSim}. A native resolution version is available online at \url{https://www.nicoledrakos.com/dream}.}
	\label{fig:wfirst_image_noisefree}
\end{figure*}

We create the synthetic images using \textsc{GalSim} \citep{galsim}. \textsc{GalSim} contains a module specifically for \emph{Roman} observations \citep[][]{troxel2021}, which includes five of the \emph{Roman} filters: Z087, Y106, J129, H158 and F184. We model each galaxy as a Sérsic profile, with an index $n_s$, axis ratio $q$ and PA, as described in Section~\ref{sec:morph}. We truncate the distribution of Sérsic indices between 0.3 and 6.2, to avoid numerical inaccuracies at more extreme values. The scale radius of the Sérsic profile, $r_0$, is directly related to the assigned half-light radius, $R_{\rm eff}$:
\begin{equation}
r_0 =  \dfrac{R_{\rm eff}}{b^{n_s} \sqrt{q}} \,\,\, ,
\end{equation}
where $b\approx 2n_s-1/3$ \citep{moriondo1998}. Fig.~\ref{fig:wfirst_image_noisefree} shows a composite of the  full synthetic galaxy catalog using Z087 (blue), Y106 (green), and H158 (red) filters. We note that we do not include simulated stars in any of the released images.

To simulate a \emph{Roman} $30~m_{\rm AB}$ survey, we convolve the image with the \emph{Roman}  point spread function (PSF), and add noise to the image. We calculate the image noise assuming proposed exposure times from \cite{koekemoer2019}, to reach a $5\sigma$ limit of $m_{\rm AB} \approx 30$ (see Table~\ref{tab:exp}). We add noise to the image in the same manner as \cite{troxel2021}. First, we generate the sky background accounting for the stray light and thermal emission from the telescope which is added to the image. Given this, we add errors associated with poisson noise, reciprocity failure, dark current, the calibration of the Roman detectors, inter-pixel capacitance, and instrument read noise. 

\begin{table}
   \caption{Approximate exposure times for the simulated filters, as calculated in \cite{koekemoer2019} to reach a $5\sigma$ limit of $m_{\rm AB} \approx 30$.}  
\centering
 \label{tab:exp}
  \begin{tabular}{c c} 
  \tableline
    Filter & Exposure Time (hrs)\\
 \tableline
    Z087 & 60  \\
   Y106 & 70 \\
   J129 &90 \\
    H158 & 40  \\
    F184 & 60   \\
 \tableline
  \end{tabular}
\end{table}

\begin{figure*}
	\includegraphics[]{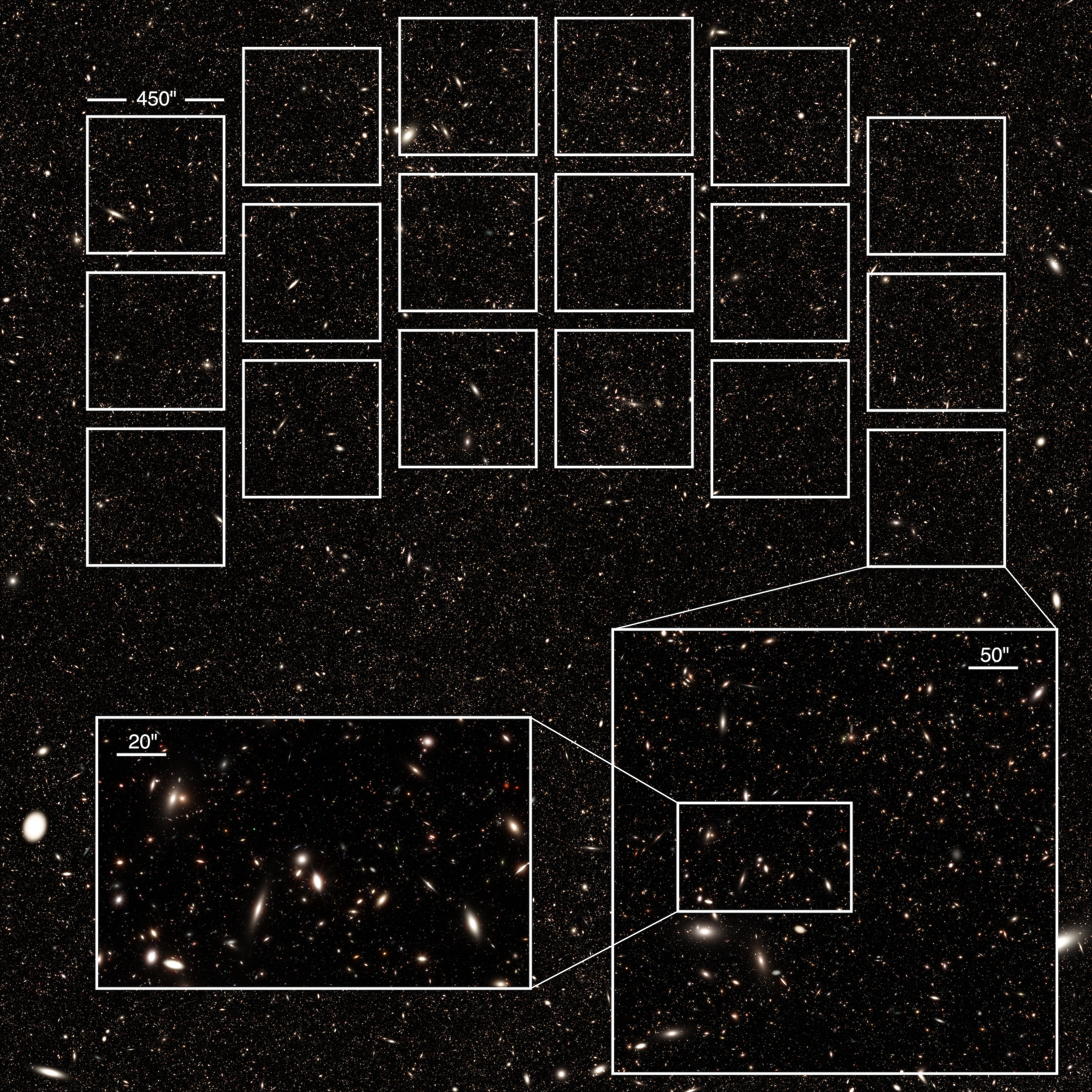}
	\caption{Composite simulated image of a region of the galaxy catalog using the Z087 (blue), Y106 (green), and H158 (red) filters convolved with the \emph{Roman} PSF. Noise is calculated assuming a depth of $\sim30\,m_{\rm AB}$ in each filter. The image was created with the \emph{Roman} module in \textsc{GalSim}. The \emph{Roman} footprint is shown on top, as well insets showing a zoomed in region of the image. A native resolution version of this $1\deg^2$ visualization is available online at \url{https://www.nicoledrakos.com/dream}.}
	\label{fig:image_composite}
\end{figure*}

We present a RGB visualization of the full \emph{Roman} catalog, with the PSF and noise included in Fig~\ref{fig:image_composite},  with one \emph{Roman} footprint overlaid on top. Each \emph{Roman} pointing will have 18 detectors.  This visualization shows the rich amount of structure that will be contained in a \emph{Roman} UDF. As described in \cite{koekemoer2019}, a $1 \deg^2$ UDF would consist of three \emph{Roman} pointings.  Though a UDF based on three tiled WFI pointings would not be perfectly square, and the exposure time would not be perfectly uniform, these synthetic images demonstrate the richness of a  $1 \deg^2$ UDF and will be incredibly useful in designing and preparing for wide, deep surveys.

\begin{figure*}
	\includegraphics[]{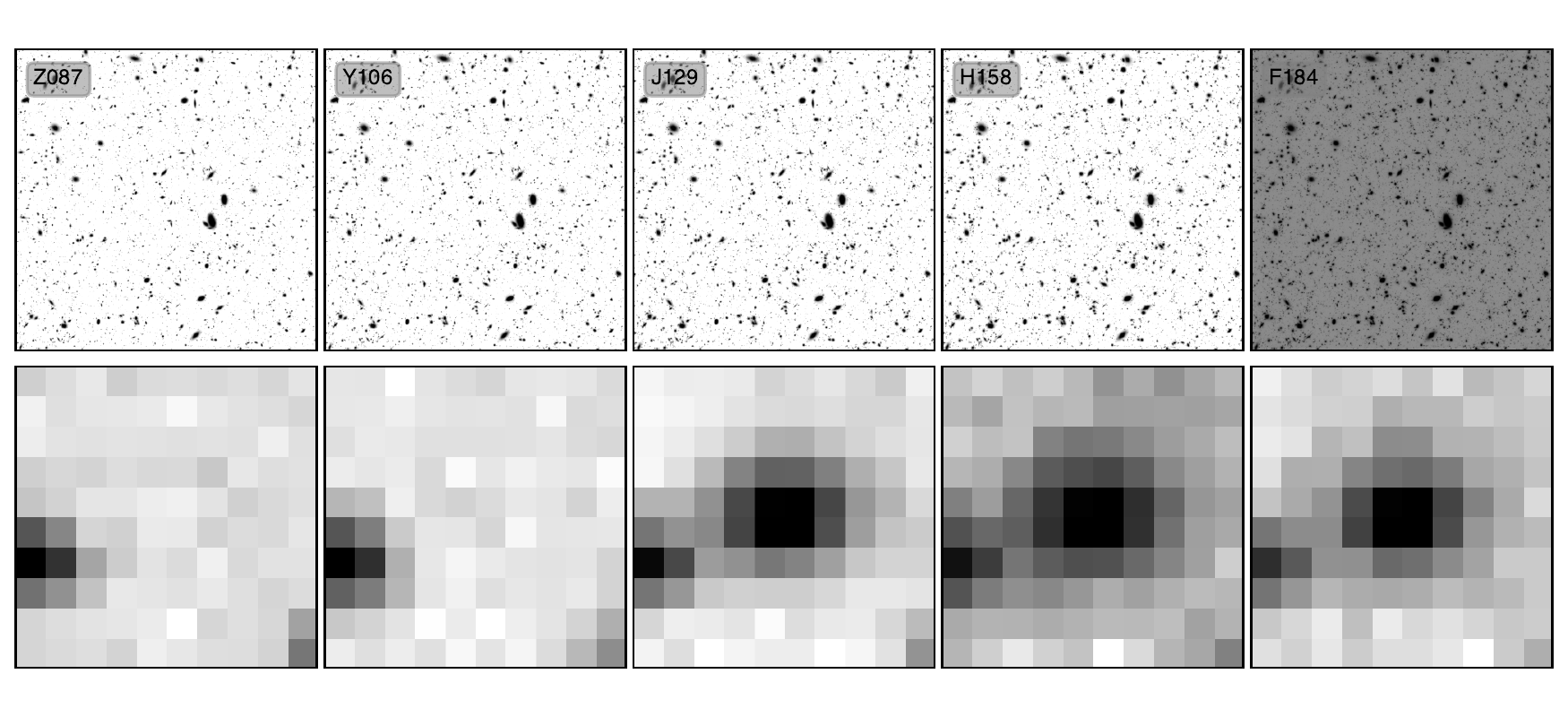}
	\caption{Flux in one WFI detector (top row), and an example of a Y106 dropout galaxy (bottom row). Each column corresponds to a different photometric filter, as labeled. The WFI detector has 4kx4k pixels, with an angular resolution of $0.11 \arcsec$ per pixel. The example dropout galaxy is at redshift $z=9.4$, has a galaxy mass of $M_{\rm gal} = 10^{7.6}\, M_{\odot}$.} 
\label{fig:fluxes_panels}
\end{figure*}

We include FITS maps of the galaxy catalog for all 5 filters currently included in GalSim in our data release. The top row of Fig.~\ref{fig:fluxes_panels} shows the flux in one WFI detector (each detector has an area of $\sim7.3\,\arcmin \times 7.3\,\arcmin $). The bottom row shows an example $z=9.4$ galaxy (with the background subtracted in each panel). This galaxy would be selected as a Y106 dropout. The dropout galaxy has a size  of $R_{\rm eff} = 0.38$\,kpc, and therefore is not resolved with the WFI (which has an angular resolution of $0.11 \arcsec$ per pixel).

The synthetic images we release are at the resolution of the WFI detector, $0.11 \arcsec$ per pixel. An actual \emph{Roman} UDF would likely be dithered, which would sample the sky on a sub-sampled pixel scale, which would improve the PSF sampling and angular resolution. Since this process would create correlations between adjacent pixels, we did not include this in our data release. Future work will build upon DREaM to examine dithering strategies of a \emph{Roman} UDF.

\section{Discussion} \label{sec:discuss}

This work presents synthetic galaxy catalogs for a $1 \deg^2$  UDF with \emph{Roman}, created using DREaM, a model for deep, realistic realizations of galaxy populations. Our model successfully reproduces a number of well-known trends, including the size--mass relation, the fundamental metallicity relation, and UV luminosity functions. Additionally, we reproduce observed low-redshift galaxy clustering and SMFs. We have made the galaxy catalog, and synthetic images public, to provide the community tools to prepare for a potential UDF with \emph{Roman}.

A UDF survey with  \emph{Roman} will address a number of science topics, including the EoR, the emergence of quiescent galaxies, and the high-redshift galaxy--halo connection. \emph{Roman}'s large field of view will capture enormous sample sizes, over contiguous fields, imaging multiple ionization bubbles. This large area will decrease cosmic variance, and probe detailed environments around high-redshift galaxies.

\subsection{The epoch of reionization}

The EoR is the final frontier for galaxy surveys.  Given the difficulty in measuring galaxies at high redshifts, this period in the universe's history is remarkably unconstrained. High-redshift low-mass galaxies were likely the major source of the ionizing photons in the EoR \citep[e.g.][]{finkelstein2019}, and observations indicate that reionization was a ``patchy'' process \cite[e.g.][]{furlanetto2005, villasenor2021}. To fully understand the EoR, we need a complete census of galaxies and their ionizing photon contribution.

To determine whether there are enough high-redshift, faint galaxies to cause reionization, we must accurately model the CSFRD, which depends on the UVLF. Though there are some constraints on the UVLF to  $z\sim10$ \citep[e.g.][]{bouwens2021}, there is still much improvement to be made on constraining both the bright steep end of the UVLF and the faint end  \citep[e.g.][]{bowler2014,bowler2015} at high redshifts. \emph{JWST} will make improvements on this front. For example, \cite{kauffmann2020} predicts that the $100\,{\rm arcmin}^2$ Cosmic Evolution Early Release Science Survey  (CEERS), will constrain the faint-end of the UVLF to a precision of $0.25$ at $z\ge8$, but that a survey would  need to be at least $300\,{\rm arcmin}^2$ to constrain bright end up to $z=8$. We predict a \emph{Roman} UDF will capture enough high-redshift galaxies  to constrain the UVLF to within 1\% at redshifts beyond $z\sim10$. This will either confirm that there are enough faint galaxies to account for the ionizing photons needed to cause reionization, or indicate that another source, such as AGN \citep[e.g.][]{madau2015} must contribute.

In addition to the abundance of galaxies, the total ionizing photon budget depends on the Lyman-continuum (LyC) production efficiency, $\xi_{\rm ion}$, and the LyC escape fraction,  $f_{\rm esc}$. To account for reionization, galaxies need higher escape fractions at high redshifts than have been observed at low redshifts \citep{davies2021}. Since the neutral IGM absorbs LyC photons, $f_{\rm esc}$ is very difficult to constrain at the EoR. However, $f_{\rm esc}$ could possibly be measured during the EoR using indirect methods \cite[e.g.][]{leethochawalit2016,zackrisson2017,chisholm2018,chisholm2020}. 
A \emph{Roman} UDF will detect tens of thousands of galaxies during the epoch of reionization, including rare, bright sources that are ideal for spectroscopic follow-up to measure $f_{\rm esc}$. Further, since \emph{Roman} can map out the density around each galaxy, it will allow for the measurement of the escape fraction in different environments.  

\subsection{The emergence of quiescent galaxies} \label{sec:discussQGs}

There exists a clear bimodality in star formation rates, indicating two distinct populations (star-forming and quiescent). The decline in the CSFRD since redshift $z\approx2$ was likely caused by the quenching of galaxies \citep[e.g.][]{renzini2016}. However, the mechanisms that transforms galaxies from star-forming to quiescent are not fully understood. Possible mechanisms that have been proposed include feedback from stars and AGN, or the removal of gas through tidal or ram pressure stripping. Quenching is likely caused by a combination of these mechanisms, and the dominant processes may be redshift dependant \citep[][]{kalita2021}.

Advances in the understanding of the origin of quiescent galaxies is being greatly improved by surveys such as the spectroscopic survey Gemini Observations of Galaxies in Rich Early ENvironments \cite[GOGREEN;][]{balogh2021}, which targets quiescent galaxies in clusters around redshift $z \sim 1$. \emph{Roman} has the potential to identify quiescent galaxy populations out to much higher redshifts. In particular, the synthetic catalog presented in this work predicts that an UDF with \emph{Roman} will contain  $\sim 10^5$ detectable quiescent galaxies, including QGs beyond redshift $z=7$. Though the exact number of quiescent galaxies relies heavily on the underlying assumed SMF, \emph{Roman} will likely detect the highest redshift QG to date if they exist, allowing for the study of quenching mechanisms at high redshifts.

\subsection{The galaxy--halo connection}

A \emph{Roman} UDF will likely provide the first strong clustering measurements of the faint, high-redshift galaxies responsible for reionization, which will enable a measurement the underlying dark matter mass. 
In our preliminary predictions, we have shown that the coverage from a UDF will constrain the 2PCF for faint galaxies out to redshifts $z=10$ (at the 10\% level). These measurements will likely provide the first direct measurement of halo masses at $z=10$, placing constraints the galaxy--halo connection at high redshift.  Since the galaxy--halo connection depends on both galaxy formation physics and the underlying cosmological model, this will provide important tests for current models of galaxy formation.

\subsection{Synergies with JWST} \label{sec:jwst}

\emph{Roman}'s extensive field of view will allow for the contiguous, deep imaging of galaxies, reducing cosmic variance, and probe the environment around individual galaxies. The large number of galaxies detected by a \emph{Roman} UDF will reduce poisson noise, and increase the number of rare objects that will be detected. Since a single \emph{Roman} pointing is wide enough to capture several reionization bubbles at the height of reionization, a \emph{Roman} UDF will allow for the study of differences between galaxies in ionized and neutral regions. For instance, there may be possible variations in the faint-end slope of the UVLF with environment, which can be measured with a deep, wide, galaxy survey. 

Before the launch of \emph{Roman}, \emph{JWST} will begin to address questions regarding the EoR, galaxy formation and the galaxy--halo connection. In particular, the JWST Advanced Deep Extragalactic Survey (JADES) is a $236 {\rm arcmin}^2$ planned imaging and spectroscopy survey\footnote{Other JWST programs, such as
Public Release IMaging for Extragalactic Research (PRIMER), and the COSMOS-Webb survey will cover larger areas than JADES, and probe the EoR, but neither will go as deep or as wide as a $1\deg^2$\emph{Roman} UDF.} Though JADES will provide revolutionary data, it will detect less than $\sim10\times$  the number of objects at all redshifts compared to a \emph{Roman} UDF. In addition, cosmic variance may dominate over Poisson noise for future high-$z$ surveys \citep{trapp2020}, and JADES will cover a much smaller area.

Though \emph{Roman} will cover a huge area on the sky, it will be difficult to obtain photometry at wavelengths greater than 2 microns. At high redshifts, accurate measurements of stellar masses and QG identification require supplementary imaging from another observing facility.  \emph{JWST} will probe very far into the infrared, which will potentially allow for Balmer-break selection of galaxies, improving redshift measurements and stellar mass estimates (though JWST is not sensitive to light blueward of 1 micron). Therefore, JWST may provide valuable spectroscopic followup to rare detections from a \emph{Roman} UDF. For example, \emph{Roman} will detect many  galaxies, on the bright end of the UVLF. Bright galaxies (e.g. starburst galaxies, with $M_{\rm UV}<-22$) are ideal for spectroscopic follow-up with JWST, and can be used to observe nebular lines and estimate $f_{\rm esc}$.

\subsection{Applications of synthetic images} \label{sec:discuss_img}

In addition to the simulated galaxy catalog, we provide synthetic images of a \emph{Roman} UDF. These images can be used to determine the impact of source blending,  line confusion, potential problems with SED fitting \citep[e.g.][]{borlaff2019, massara2020, kauffmann2020}. Having very deep synthetic images will be useful for studying WFI systematics, processing issues (e.g. low surface brightness issues), and secondary analysis (e.g. photo-z studies). Quantifying these issues will also be beneficial for many other \emph{Roman} surveys, such as the High Latitude Survey (HLS). For instance, in Section~\ref{sec:detect}, we assumed that all galaxies that are detectable as will be selected. However, since a UDF will be so richly populated with structure, a fraction of the high redshift  galaxies will be obscured by forefront galaxies. We intend to  use the synthetic images and catalogs in this work to quantify this effect.

\subsection{Grism predictions}

In this paper we have only included photometric predictions, but the realistically modeled galaxy SEDs can also be used to generate grism predictions. In particular, the MOSFIRE Deep Evolution Field (MOSDEF) survey \citep{kriek2015} measured  the detailed rest-frame optical emission-line SEDs in galaxies $z=2$--$3$. More recent measurements with Keck/MOSFIRE have presented spectroscopic measurements out to $z \sim 8$ \citep{topping2021a}. We can combine these observations with our synthetic catalog, to guide the \emph{Roman} grism data reduction pipeline at high redshifts.

The synthetic catalog can also be used to study Lyman-$\alpha$ emitters (LAEs). LAEs produce a large amount of ionizing photons, and both UVLFs and clustering of LAEs are important to characterize EoR. The Ly$\alpha$ UVLF decreases towards early stage of EoR, since neutral hydrogen absorbs Lyman-$\alpha$ photons \citep[e.g.][]{ouchi2018}.  The clustering of LAEs emitters constrains the ionized fraction and topology of reionization.

\section{Summary and Conclusions}

This work presents DREaM, a Deep Realistic Extragalactic Model for creating synthetic galaxy catalogs. We use this model to understand the potential power of a $1 \deg^2$  UDF with  \emph{Roman}, and provide publicly available realistic synthetic galaxy catalogs and images. The synthetic catalogs and images will aid the community in designing and interpreting a \emph{Roman} UDF.   A summary of our main predictions are below.

A $1 \deg^2$ \emph{Roman} UDF will:

\begin{enumerate}

\item{contain more than $10^6$ detectable galaxies, with more than $10^4$ during the EoR ($z>7$).}

\item{contain $\sim10^5$ detectable quiescent galaxies, including a few at redshifts beyond $z\sim7$, likely detecting the farthest redshift quiescent galaxy to date.}

\item{help constrain SMFs for SFGs to redshifts beyond $z\sim10$, and for QGs past redshift $z\sim7$.}

\item{provide tight constraints (within 1\% percent) on the faint-end  ($M_{\rm UV}<-17$) of the UV luminosity function, out to redshifts $z\sim10$.}

\item{provide high redshift ($z>7$) constraints on the clustering of the faint galaxies thought to be responsible for reionization.}

\item{look for variations in the UVLF in different environments.}

\end{enumerate}

Overall, \emph{Roman}'s wide field of view offers a unique ability to create wide, deep surveys.  A \emph{Roman} UDF would enable a tremendous amount of science by detecting the largest census of high-redshift galaxies to date, and differentiating between galaxy populations in low and high density regions during the EoR.

\acknowledgments

This work was supported by NASA contract NNG16PJ25C. The authors acknowledge use of the lux supercomputer at UC Santa Cruz, funded by NSF MRI grant AST 1828315, and the NASA supercomputers. In addition, the authors thank Christina Williams, Yifei Luo, David O. Jones and Kevin Hainline for useful discussion. 

\emph{Software}: numpy \citep{numpy}, matplotlib \citep{matplotlib},  scipy \citep{scipy}, astropy \citep{astropy1, astropy2}, python-fsps \citep{pythonfsps}

\clearpage
\appendix

\section{Data Release}\label{sec:datarel}

We have made a number of our data products public, as summarized in Table~\ref{tab:DataProd}. We release a main galaxy catalog, a catalog containing intrinsic galaxy properties, and a halo catalog. The galaxy properties that are included in the catalogs are summarized in Table~\ref{tab:CatContent}. These data products, and an interactive online visualization of the synthetic images are available at \url{https://www.nicoledrakos.com/dream}.

\begin{table*}
  \centering
   \caption{Data Products. Available online at \url{https://www.nicoledrakos.com/dream}.}  \label{tab:DataProd}
  \begin{tabular}{p{0.2\linewidth} p{0.2\linewidth} p{0.5\linewidth}} 
  \tableline
     Product &  Filename &Description\\
 \tableline
   Main Catalog 
	& \texttt{DREaM\_main.fits}
	& Contains galaxy masses, positions, morphologies, rest-frame properties, and photometry in \emph{Roman} and \emph{JWST} filters. See Table~\ref{tab:CatContent} for more information.\\ \hline
Intrinsic Properties
	& \texttt{DREaM\_intrinsic.fits}
	& Contains the FSPS parameters used to generate galaxy SEDs. See Table~\ref{tab:CatContent} for more information. \\ \hline
Halo Properties
	& \texttt{DREaM\_halos.fits}
	&Contains the host halo properties, such as the mass, shape, size, spin and peculiar velocity of the host halos. See Table~\ref{tab:CatContent} for more information. \\ \hline
 \tableline
  \end{tabular}
\end{table*}

\begin{table*}
  \centering
   \caption{Catalog content. The galaxy information is contained in either the main catalog (MC), internal properties catalog (IPC) or halo properties catalog (HPC).}  \label{tab:CatContent}
  \begin{tabular}{p{0.35\linewidth}  p{0.46\linewidth} p{0.03\linewidth} p{0.03\linewidth} p{0.03\linewidth}} 
  \tableline
Variables &  Description & MC &IPC &HPC\\
 \tableline
 ID &  Galaxy ID. The same across all the catalogs &\cmark &\cmark &\cmark \\
 RA, Dec &   Right Ascension and Declination  [degrees, $-0.5$ to $0.5$]&\cmark &\cmark &\cmark \\
redshift & Galaxy redshift &\cmark &\cmark &\cmark \\
M\_halo & Halo mass [$M_\odot/h$] &\cmark &\cmark &\cmark \\
M\_gal  & Galaxy mass  [$M_\odot$] &\cmark &\cmark &\cmark \\
logpsi  & Star-formation rate [$\log_{10}(\psi/(M_\odot/ {\rm yr} ))$] &\cmark &\cmark &\cmark \\
\hline
M\_UV  & Rest-frame UV magnitude&\cmark &  \xmark& \xmark \\
R\_eff  &  Half-light radius in the semi-major axis [kpc/h, physical] &\cmark & \xmark& \xmark \\
n\_s  & Sérsic index&\cmark &\xmark& \xmark \\
q  &  Projected axis ratio: semi-minor to semi-major half-light size &\cmark & \xmark& \xmark \\
PA   &Position angle [radians, $0$ to $2\pi$] &\cmark &\xmark& \xmark \\
beta  & Rest-frame UV continuum slope &\cmark & \xmark& \xmark \\
U, V, J  & Rest-frame magnitude in U, V and J bands&\cmark & \xmark& \xmark \\
R062,  Z087, Y106, J129, H158, F184, F213 & Apparent AB magnitude in \emph{Roman} filters&\cmark & \xmark& \xmark \\
f070, f090, f115, f150, f200, f277, f356, f444&Apparent AB magnitude in \emph{JWST} filters&\cmark & \xmark& \xmark \\
\hline
SF & Star-forming (True) or Quiescent (False) & \xmark &\cmark& \xmark \\
t\_start & Age of universe when galaxy started forming [Gyr]& \xmark &\cmark& \xmark \\
tau & $e$-folding time for star formation [Gyr]&\xmark &\cmark& \xmark \\
logZ & Metallicity parameter [$\log_{10}(Z/Z_\odot)$]&\xmark &\cmark& \xmark \\
dust & Dust attenuation parameter& \xmark &\cmark& \xmark \\
logUS & Gas ionization parameter [$\log_{10}(U_s)$]& \xmark &\cmark& \xmark \\
\hline
V\_max & Halo maximum circular velocity [km/s, physical]& \xmark &\xmark& \cmark\\
R\_s & Halo scale radius [kpc/$h$, comoving]&\xmark &\xmark& \cmark\\
R\_vir & Halo virial radius [kpc/$h$, comoving]& \xmark &\xmark& \cmark\\
M\_peak &  Halo peak mass over accretion history [$M_\odot/h$]& \xmark &\xmark& \cmark\\
V\_peak & Halo peak $V_{\rm max}$ over accretion history [km/s, physical]&\xmark &\cmark& \xmark\\
b\_a & Halo axis ratio, $b/a$&\xmark &\xmark& \cmark\\
c\_a & Halo axis ratio, $c/a$& \xmark &\xmark& \cmark\\
haloID & ID of host halo&\xmark &\xmark& \cmark\\
hostID & ID of least massive host halo (-1 if distinct halo).& \xmark &\xmark& \cmark\\
spin\_B & \cite{bullock2001b} halo spin parameter & \xmark &\xmark& \cmark\\spin\_P & \cite{peebles1971} halo spin parameter & \xmark &\xmark& \cmark\\
x,y,z & Halo positions [Mpc/$h$, comoving] & \xmark &\xmark& \cmark\\
vx,vy,vz &Halo peculiar velocities [km/s, physical] &\xmark &\xmark& \cmark\\
 \tableline
  \end{tabular}
\end{table*}

\section{Synthetic galaxy properties}\label{sec:galprops}

Our  main goal was to  study the ability of a \emph{Roman} UDF to constrain the photoionizing contribution of high redshift galaxies, and the environments around these galaxies. Therefore, we carefully constructed the DREaM galaxy catalogs to have realistic clustering properties and UV properties. In addition to these properties, our synthetic galaxies also capture a number of other observational trends.

In this section we examine some of the properties of the full galaxy catalog. We begin by looking at the how the UVJ colors (Section~\ref{sec:QG_props}), ages (Section~\ref{sec:age})  and SFRs (Section~\ref{sec:SFR}) differ between star-forming and quiescent galaxies. Additionally we verify that we reproduce the well-known fundamental metallicity relation (Section~\ref{sec:metal}).

\subsection{UVJ colors} \label{sec:QG_props}

\begin{figure*}
	\includegraphics[]{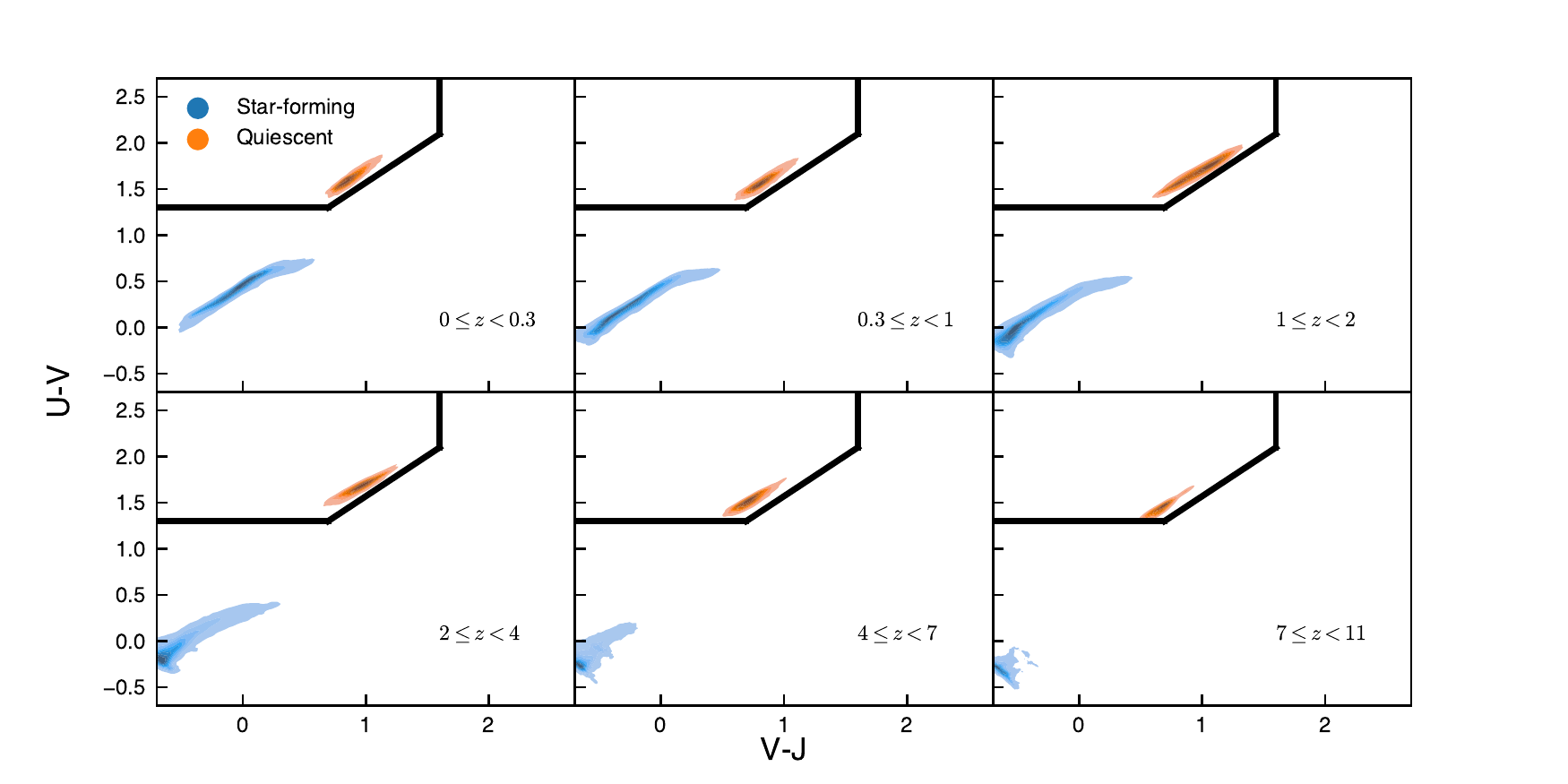}
	\caption{UVJ diagram of synthetic SFGs (blue)  and QGs (orange). We used the selection box from \cite{williams2009} (black lines). Our synthetic galaxies demonstrate a clear bimodal distribution, with SFGs falling in the top left corner of the UVJ diagram.}
	\label{fig:UVJ}
\end{figure*}

As discussed in Section~\ref{sec:UVJ}, UVJ diagrams differentiate between star-forming and quiescent galaxy populations. We display the U-V and V-J colors for our two galaxy populations in Fig.~\ref{fig:UVJ}, along with the UVJ selection box from \cite{williams2009}. The two populations occupy two distinct regions in this parameter space, demonstrating that the synthetic galaxy catalog does capture the bimodal population. Observational constraints for QGs do not currently exist for galaxies past redshifts $z\approx 4$, the UVJ distribution of the synthetic galaxies at low redshifts do agree with observations \citep[e.g.][]{schreiber2015}.

\subsection{Star formation rate} \label{sec:SFR}

We do not explicitly assign SFRs to the galaxies in the mock catalog, but we reproduce realistic trends in SFRs. By assigning realistic $M_{\rm UV}$ values to the SFGs, and constraining QGs to the appropriate place in the UVJ color diagrams, we accurately model a bimodal population, with SFGs having higher SFRs than QGs. Additionally we reproduce the observed cosmic star formation rate density (CSFRD) as demonstrated in Section~\ref{sec:csfrd}. 

\begin{figure}
	\includegraphics[width=\columnwidth]{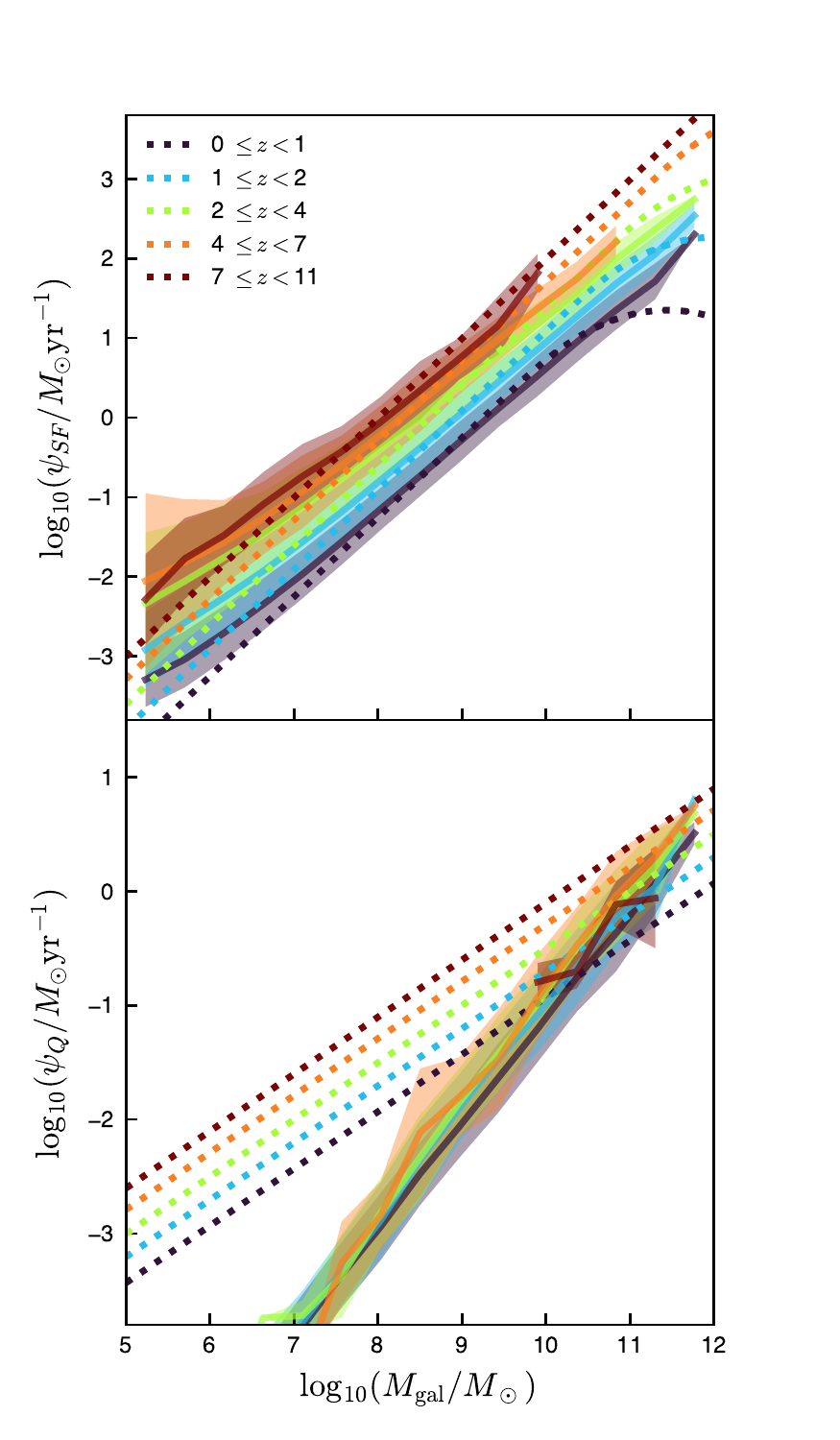}
	\caption{The average SFR versus galaxy mass for synthetic SFGs (top) and QGs (bottom).  The SFRs increase with mass, and the SFGs have higher SFRs than QGs, as expected. For comparison, we plot the relation from \cite{schreiber2017} (dotted lines). The galaxy catalog agrees closely with the \cite{schreiber2017} parameterization for the SFGs. The QGs have lower SFRs than the \cite{schreiber2017} parameterization, but are still consistent with observations (see discussion in text).}
	\label{fig:SFR_vs_M}
\end{figure}

Fig.~\ref{fig:SFR_vs_M} shows the SFR--Mass relation of the synthetic catalog compared to the relations from \cite{schreiber2017}. We closely match the \cite{schreiber2017} SFG SFR--mass relations. The SFRs for QGs is lower than the \cite{schreiber2017} relation. However, the SFRs--mass relations from \cite{schreiber2015} assumed that all IR emission from QGs originated from residual star-formation. Alternative explanation are that the IR emission originates from active galactic nuclei (AGN) torus emission, dust heating, or incorrect classification of SFGs. The QGs SFRs still agree with what current observations can predict.

\subsection{Age}\label{sec:age}

Galaxy ages are a measure of star-formation histories (SFHs). SFHs of individual galaxies, give a direct measurement of the evolution of different galaxy populations. Trends in galaxy ages have been well-established; for instance, QGs are older than SFGs, and high-redshift galaxies are older than low-redshift galaxies \citep[e.g.][]{webb2020}. However, galaxy ages for individual galaxies are difficult to measure due to degeneracies between other parameters (metallicity, dust), sensitivity to priors in fitting \citep[e.g.][]{leja2019a}, and the similarity between SEDs in galaxies older than $\sim 5$ Gyr \citep[e.g.][]{gallazzi2005}. Our synthetic galaxy catalogs provide age estimates, and realistic SEDs to examine this further.

We calculate the mass-weighted age of the synthetic galaxies as 
\begin{equation} \label{eq:agal}
a_{\rm gal } = \dfrac{\int_0^{t_{\rm age}} (t_{\rm age} - t) \psi dt  }{\int_0^{t_{\rm age}}  \psi dt} \,\,\, .
\end{equation}
For the delayed-tau model, 
\begin{equation}
a_{\rm gal } = t_{\rm tot} - 2 \tau + \dfrac{t_{\rm tot}^2}{\tau e^{t_{\rm tot} /\tau} - (\tau+t_{\rm tot} )} \,\,\, ,
\end{equation}
where $t_{\rm tot}=t_{\rm age} - t_{\rm start} $ is the total time of star formation. 

\begin{figure}
	\includegraphics[width=\columnwidth]{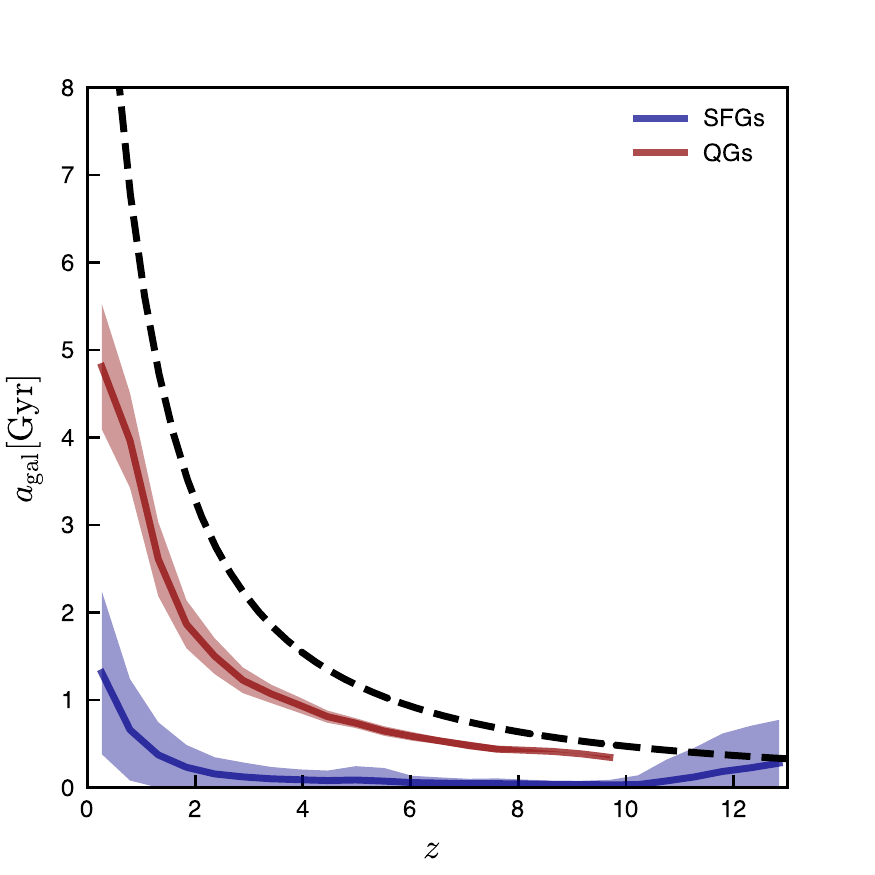}
	\caption{Average mass-weighted age of synthetic galaxies for SFGs (blue points) and QGs (red points), as defined in Equation~\eqref{eq:agal}. QGS are older than SFGs, and ages decrease with increasing redshift.  Galaxy ages are less than the age of the universe (dashed black line).}
	\label{fig:age_zred}
\end{figure}

Fig.~\ref{fig:age_zred} shows the average galaxy ages as a function of redshift  for both SFGs (blue) and QGs (red). QGs are older than SFGs at all redshifts and age decreases with increasing redshift, as expected. At low redshifts, QGs are $\approx 3$\,Gyr older than SFGs.

\subsection{Metallicity} \label{sec:metal}

\begin{figure} 
	\includegraphics[width=\columnwidth]{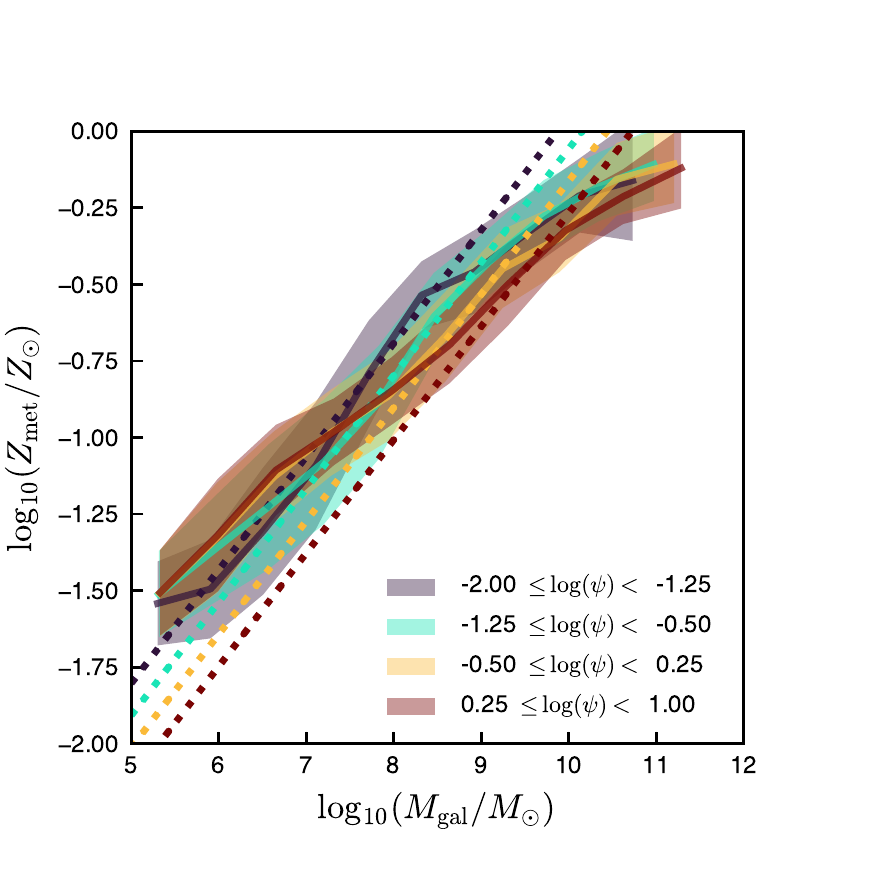}
	\caption{Mass–metallicity relation of the synthetic galaxies (solid lines). The shaded regions are one standard deviation. Dotted lines show the FMR from \cite{williams2018}, which were used to generate the parent catalog SEDs. The synthetic galaxies have increased metallicity with mass and SFR, in agreement with the underlying FMR.}
	\label{fig:FunMet}
\end{figure}

Galaxy metallicity can greatly affect quantities such as galaxy color, and therefore must be accurately modeled. We assigned metallicities from the FMR (Equation~\ref{eq:Zmet}, to the parent catalogs used in the SED pipeline. 
Fig.~\ref{fig:FunMet} shows the FMR for the synthetic galaxies. We show for comparison the FMR (dotted lines), where we have used the assumption $12 + \log_{10}({\rm O/H}) \approx \log_{10}(Z_{\rm met}/Z_{\rm sol})$.

Our synthetic galaxies follow the expected FMR, where metallicity increases with mass, with a turn-over at high masses, as seen in observations (see discussion in Section~\ref{sec:parent}). The low-mass galaxies have lower metallicities than the high-mass galaxies as expected, but do not vary much with SFR. However, at these low masses, very few galaxies will have SFRs $\psi > -1$.

\clearpage
\bibliography{MockCatBib}{}

\bibliographystyle{aasjournal}



\end{document}